\documentclass[twocolumn]{aastex63}

\usepackage{my_astro}
\usepackage{courier}
\usepackage{amsmath}
\usepackage[super]{nth}
\usepackage[color=yellow]{pdfcomment}

\newcommand{\md}{M~dwarf}

\newcommand{\teff}{$T_\mathrm{eff}$}
\newcommand{\gaia}{\textit{Gaia}}
\newcommand{\bprp}{$BP - RP$}
\newcommand{\Prot}{$P_\mathrm{rot}$}

\newcommand{\tauc}{$\tau_c$}
\newcommand{\Rhk}{$R'_{HK}$}
\newcommand{\fiteqn}{\begin{equation}
F = \left\{\begin{matrix}F_\mathrm{sat}, & x \leq x_\mathrm{sat} \\F_\mathrm{sat} \left( x/x_\mathrm{sat}\right)^{-\alpha}, & x >  x_\mathrm{sat}\end{matrix}\right.
\end{equation}}

\newcommand{\evolCaption}{
The component fluxes of multiplets have been added\deleted{, except for \Cii\ (Section \ref{subsec:lines})}.
Error bars are often smaller than the points.
Broken power law fits to the evolution of each line are overplotted with gray 68\% confidence intervals.
The y-axis of each subplot spans the same number of decades to facilitate the comparison of slopes.}

\begin{document}

\title{HAZMAT. VII. The Evolution of Ultraviolet Emission with Age and Rotation for Early M Dwarf Stars}

\correspondingauthor{R. O. Parke Loyd}
\email{parke@asu.edu}

\author[0000-0001-5646-6668]{R. O. Parke Loyd}
\affiliation{School of Earth and Space Exploration, Arizona State University, Tempe, AZ 85287}

\author[0000-0002-7260-5821]{Evgenya L. Shkolnik}
\affiliation{School of Earth and Space Exploration, Arizona State University, Tempe, AZ 85287}

\author[0000-0002-6294-5937]{Adam C. Schneider}
\affiliation{U.S. Naval Observatory, Flagstaff Station, 10391 West Naval Observatory Road, Flagstaff, AZ, 86005-8521, USA}
\affiliation{Department of Physics and Astronomy, George Mason University, 4400 University Drive, MSN 3F3, Fairfax, VA 22030, USA}

\author[0000-0003-1290-3621]{Tyler Richey-Yowell}
\affiliation{School of Earth and Space Exploration, Arizona State University, Tempe, AZ 85287}

\author[0000-0002-1386-1710]{James A. G. Jackman}
\affiliation{School of Earth and Space Exploration, Arizona State University, Tempe, AZ 85287}

\author[0000-0002-1046-025X]{Sarah Peacock}
\affiliation{Lunar and Planetary Laboratory, University of Arizona, Tucson, AZ 85721 USA}
\affiliation{NASA Goddard Space Flight Center, Greenbelt, MD 20771, USA}

\author[0000-0002-7129-3002]{Travis S. Barman}
\affiliation{Lunar and Planetary Laboratory, University of Arizona, Tucson, AZ 85721 USA}

 \author[0000-0001-9573-4928]{Isabella Pagano}
\affiliation{INAF - Osservatorio Astrofisico di Catania, Via S. Sofia 78, 95123, Catania, Italy}

\author[0000-0002-1386-1710]{Victoria S. Meadows}
\affiliation{NASA Astrobiology Institute Alternative Earths and Virtual Planetary Laboratory Teams}
\affiliation{Department of Astronomy, University of Washington, Seattle, WA, USA}

\received{2020 September 9}
\revised{2020 October 22}
\accepted{2020 November 19}
\submitjournal{The Astronomical Journal}

\begin{abstract}
The ultraviolet (UV) emission from the most numerous stars in the universe, M dwarfs, impacts the formation, chemistry, atmospheric stability, and surface habitability of their planets.
We have analyzed the spectral evolution of UV emission from M0-M2.5 (0.3-0.6~\Msun) stars as a function of age, rotation, and Rossby number, using \textit{Hubble Space Telescope} observations of Tucana Horologium (40~Myr), Hyades (650~Myr), and field (2-9 Gyr) objects.
The quiescent surface flux of their \Cii, \Ciii, \Civ, \Heii, \Nv, \Siiii, and \Siiv\ emission lines, formed in the stellar transition region, remains elevated at a constant level for \replaced{$240^{+20}_{-30}$}{$240\pm30$}~Myr before declining by 2.1 orders of magnitude to an age of 10 Gyr.
\Mgii\ and far-UV pseudocontinuum emission, formed in the stellar chromosphere, exhibit more gradual evolution with age, declining by 1.3 and 1.7 orders of magnitude, respectively.
The youngest stars exhibit a scatter of 0.1~dex in far-UV line and pseudocontinuum flux attributable only to rotational modulation, long-term activity cycles, or an unknown source of variability.
Saturation-decay fits to these data can predict an M0-M2.5 star's quiescent emission in UV lines and the far-UV pseudocontinuum with an accuracy of roughly 0.2-0.3~dex, the most accurate means presently available.
Predictions of UV emission will be useful for studying exoplanetary atmospheric evolution, the destruction and abiotic production of biologically relevant molecules, and interpreting infrared and optical planetary spectra measured with observatories like the \textit{James Webb Space Telescope}.
\end{abstract}


\section{Introduction}
Randomly pick an Earth-sized planet in the universe and, more likely than not, it will orbit an M dwarf star.
This would also be true if one were to draw a planet only from those in the classical ``habitable zone,'' the region of planetary orbits where liquid water is possible on the surface of an Earth-like planet \citep{kasting93,kopparapu13}.
This conclusion comes from the fact that M dwarfs outnumber Sun-like stars by a factor of roughly 20 \citep{henry18} and that, on average, an M dwarf will have more Earth-sized planets than an F, G, or K star \citep{mulders18}.
The space environment of M dwarfs is much different than that of the most familiar planetary system, the Solar System.
Their fractional X-ray and ultraviolet (UV) emission exceeds the Sun's by up to two orders of magnitude \citep{loyd16}, and these stars experience a comparatively prolonged period of even more elevated emission in their youth \citep{shkolnik14}.

An M dwarf's high energy photons make up only a small fraction of its total luminosity, yet they can shape a planet's atmosphere.
UV and X-ray photons break atom-atom and atom-electron bonds, chemically modifying, heating, desiccating, and evaporating planetary atmospheres and oceans (e.g., \citealt{hu12,luger15}).
Understanding the evolution of this radiation as these stars age is necessary to understand the evolution and present state of their planets.
This includes interpreting the spectra of planets, such as those soon to be observed in the optical and infrared by the next generation of ground and space-based telescopes (\hyperlink{cite.ess}{NAS Exoplanet Science Strategy Report 2018}\footnote{\url{https://www.nap.edu/catalog/25187/exoplanet-science-strategy}}).

Individual emission lines in the UV are important to both planetary and stellar physics.
The photolysis cross sections for molecules are heavily wavelength dependent (e.g., \citealt{segura05}).
Because of this, spectrally-resolved UV input is essential to photochemical models of exoplanetary atmospheres.
UV emission lines can also be a tool to guide physical models of stellar upper atmospheres that in turn can predict important regions of stellar emission that have not or cannot be observed, such as \replaced{the extreme UV}{extreme UV (EUV) emission absorbed by intervening gas in the interstellar medium (ISM)} (\deleted{EUV; }\citealt{fontenla16,peacock20}\added{; see also Section \ref{subsec:king}}).
The utility of UV lines in constraining models comes from the fact that various lines within the UV band include significant emission from chromospheric, transition region, coronal, and even photospheric plasma.
In the same vein, UV spectra have been used in differential emission measure analyses to predict optically thin emission in the EUV (e.g., \citealt{pagano00}).

In this work, we present observations of the UV spectral evolution of early M stars\footnote{Throughout this work, we generally use the term ``M star'' as opposed to ``M dwarf'' to encompass young objects that have not yet fully contracted onto the main sequence.}  with age.
This is the latest in a body of work from the HAbitable Zones and M dwarf Activity across Time (HAZMAT) program.
Earlier work by the HAZMAT program leveraged archival data from the \textit{GALEX}  and \textit{ROSAT} missions to measure changes in the broadband near-UV (NUV), far-UV (FUV), and X-ray emission of M stars \citep{shkolnik14,schneider18}, and combined these data with sophisticated atmospheric models to reconstruct the evolution of M star EUV emission \citep{peacock20}.
The program has also investigated the time variability of UV emission on minute to year timescales \citep{miles17}, including temporally and spectrally resolved UV observations of flares on 40~Myr M stars \citep{loyd18a}.
\cite{richey19} recently extended HAZMAT into an investigation of K star activity evolution.
New \textit{Hubble Space Telescope (HST)} observations to deepen this investigation are underway (program 15955, PI Richey-Yowell).

\added{Beyond the HAZMAT program, most studies of M~dwarf activity in recent decades have primarily focused on \Ha\ and X-ray emission as a function of rotation \citep{stauffer97,delfosse98,pizzolato03,wright11,wright18b}.
The age evolution of X-ray emission has been addressed by \cite{feigelson04,preibisch05}; and \cite{jackson12}.
In the most closely related study outside of the HAZMAT program, \citet{stelzer13} \citep[erratum: ][]{stelzer14} examined the X-ray and broadband UV emission of a 10~pc sample of M~dwarfs, fitting an unbroken power law to the age evolution of their emission.}
This paper \replaced{presents data from a campaign of spectroscopic UV observations with \textit{HST} (program 14784, PI Shkolnik) for three diagnostic age groups (40~Myr, 650~Myr, and several Gyr) of M0-M2.5 stars by HAZMAT.
It provides a}{is the} first look into the age evolution of specific \added{chromospheric and transition-region} UV emission lines for early M stars.


\section{Observations and Stellar Sample}
\label{sec:obs}

The HAZMAT \textit{HST} survey (program ID 14784, PI Shkolnik) collected data using the Cosmic Origins Spectrograph (COS; \citealt{green12}) with the G130M and G160M ($R = $ 12,000-20,000) and G230L ($R = $ 2,000-4,000) gratings.
Observations were made of 21 M stars, 12 from the Tucana-Horologium (Tuc-Hor; 40~Myr; \citealt{kraus14}) moving group, 6 from the Hyades cluster (650~Myr; \citealt{martin18}), and 3 unassociated field stars (2-10~Gyr).

We incorporate some additional data from the \textit{HST} archive.
The field stars build upon the archival sample of 4 objects from the MUSCLES Treasury survey for a total of 7 field stars (Measurements of the Ultraviolet Spectral Characteristics of Low-mass Exoplanetary Systems Treasury survey; program IDs 12464 and 13650; PI France; \citealt{france16,youngblood16,loyd16}).
We also included archival data from programs 14767 (PI Sing; \citealt{dos19}) and 15174 (PI Loyd) for one of the 7 field stars, GJ~436.
Table \ref{tbl:starprops} gives parameters of the stellar sample and Table~\ref{tbl:obssum} summarizes the observations.
Note that the FUV and NUV observations, while often closely spaced in time, are not simultaneous.
We retrieved \textit{K2} and \textit{TESS} light curves for all targets \added{from the Mikulski Archive for Space Telescopes (MAST\footnote{\url{https://archive.stsci.edu}})} where available at the time of analysis to measure rotation periods.
\deleted{All data are available through the Mikulski Archive for Space Telescopes (MAST\\footnote{\url{https://archive.stsci.edu}}).}
\added{The \textit{HST} data described here may be obtained from the MAST archive through a single DOI portal at
\dataset[doi:10.17909/t9-se6p-dn04]{https://dx.doi.org/10.17909/t9-se6p-dn04}.}

\begin{deluxetable*}{lllrrrrrrrrrrrr}
\tablewidth{0pt}
\tabletypesize{\footnotesize}
\rotate
\tablecaption{Properties of the stellar sample. \label{tbl:starprops}}
\tablehead{\colhead{Group} & \colhead{Star} & \colhead{2MASS} & \colhead{Spectral} & \colhead{Ref} & \colhead{Alternate} & \colhead{$T_\mathrm{eff}$\tablenotemark{b}} & \colhead{Radius} & \colhead{Age} & \colhead{$P_\mathrm{rot}$\tablenotemark{c}} & \colhead{Ref} & \colhead{$Ro$} & \colhead{[FeH]} & \colhead{Ref} & \colhead{Distance\tablenotemark{d}}\\
\colhead{} & \colhead{} & \colhead{Designation} & \colhead{Type} & \colhead{} & \colhead{Type\tablenotemark{a}} & \colhead{K} & \colhead{$R_\odot$} & \colhead{Myr} & \colhead{d} & \colhead{} & \colhead{} & \colhead{} & \colhead{} & \colhead{pc}}
\startdata
Tuc-Hor & J03315 &  J03315564-4359135 & M0.0 $\pm$ 0.6 & 1 & K5.2 $\pm 1$ & $ 3959 \pm 80 $ & $ 0.824 \pm 0.087 $ & $ 40 \pm 5 $ & $ 2.92 \pm 0.01 $\tablenotemark{e} & 2 & 0.089 & \nodata &  & $ 45.275 \pm 0.071 $\\
  & J00240 &  J00240899-6211042 & M0.2 $\pm$ 0.9 & 1 & K7.2 $\pm 1$ & $ 3777 \pm 80 $ & $ 0.858 \pm 0.091 $ & $ 40 \pm 5 $ & $ 1.76 \pm 0.03 $\tablenotemark{e} & 3 & 0.044 & \nodata &  & $ 44.2 \pm 1.1 $\\
  & J02543 &  J02543316-5108313 & M1.4 $\pm$ 0.7 & 1 & M1.1 $\pm 1$ & $ 3563 \pm 80 $ & $ 0.828 \pm 0.088 $ & $ 40 \pm 5 $ & \nodata &  & \nodata & \nodata &  & $ 43.76 \pm 0.21 $\\
  & J23261 &  J23261069-7323498 & M1.5 $\pm$ 1 & 1 & K7.7 $\pm 1$ & $ 3762 \pm 80 $ & $ 0.740 \pm 0.078 $ & $ 40 \pm 5 $ & $ 0.57 \pm 0.01 $\tablenotemark{e} & 3 & 0.014 & \nodata &  & $ 46.294 \pm 0.059 $\\
  & J01521 &  J01521830-5950168 & M1.6 $\pm$ 0.4 & 1 & M1.6 $\pm 1$ & $ 3531 \pm 80 $ & $ 0.656 \pm 0.069 $ & $ 40 \pm 5 $ & $ 6.35 \pm 0.06 $\tablenotemark{e} & 2 & 0.12 & \nodata &  & $ 39.765 \pm 0.040 $\\
  & J00393 &  J00393579-3816584 & M1.8 $\pm$ 0.4 & 1 & M1.4 $\pm 1$ & $ 3555 \pm 80 $ & $ 0.693 \pm 0.073 $ & $ 40 \pm 5 $ & 6.32 & 4 & 0.12 & \nodata &  & $ 40.241 \pm 0.070 $\\
  & J02001 &  J02001277-0840516 & M2.0 $\pm$ 0.2 & 1 & M2.1 $\pm 1$ & $ 3480 \pm 80 $ & $ 0.662 \pm 0.070 $ & $ 40 \pm 5 $ & 1.75 & 5 & 0.029 & \nodata &  & $ 36.926 \pm 0.069 $\\
  & J02365 &  J02365171-5203036 & M2 & 6 & \nodata & $ 3504 \pm 80 $ & $ 0.809 \pm 0.086 $ & $ 40 \pm 5 $ & $ 0.74 \pm 0.01 $\tablenotemark{e} & 3 & 0.013 & \nodata &  & $ 38.847 \pm 0.051 $\\
  & J22025 &  J22025453-6440441 & M2.1 $\pm$ 0.6 & 1 & M1.8 $\pm 1$ & $ 3525 \pm 80 $ & $ 0.678 \pm 0.072 $ & $ 40 \pm 5 $ & $ 0.43 \pm 0.01 $\tablenotemark{e} & 3 & 0.0073 & \nodata &  & $ 43.705 \pm 0.098 $\\
  & J23285 &  J23285763-6802338 & M2.9 $\pm$ 0.5 & 1 & M2.3 $\pm 1$ & $ 3444 \pm 80 $ & $ 0.666 \pm 0.070 $ & $ 40 \pm 5 $ & 0.74 & 5 & 0.012 & \nodata &  & $ 46.048 \pm 0.050 $\\
  & J22463 &  J22463471-7353504 & M3.2 $\pm$ 0.3 & 1 & M2.3 $\pm 1$ & $ 3451 \pm 80 $ & $ 0.609 \pm 0.064 $ & $ 40 \pm 5 $ & $ 1.65 \pm 0.01 $\tablenotemark{e} & 3 & 0.027 & \nodata &  & $ 50.224 \pm 0.074 $\\
  & J02125 &  J02125819-5851182 & M3.5 $\pm$ 0.5 & 1 & M1.9 $\pm 1$ & $ 3513 \pm 80 $ & $ 0.671 \pm 0.071 $ & $ 40 \pm 5 $ & 1.60 & 3 & 0.028 & \nodata &  & $ 48.061 \pm 0.049 $\\
Hyades & HAN~192 &  J04184702+1321585 & M0.5 $\pm$ 0.5 & 3 & \nodata & $ 3922 \pm 88 $ & $ 0.624 \pm 0.086 $ & $ 650 \pm 70 $ & $ 14 \pm 4 $ & 3 & 0.39 & $ 0.146 \pm 0.004 $ & 7 & $ 46.14 \pm 0.16 $\\
  & TYC~1265 &  J04260470+1502288 & M1 $\pm$ 0.5 & 3 & \nodata & $ 3901 \pm 88 $ & $ 0.601 \pm 0.082 $ & $ 650 \pm 70 $ & $ 16 \pm 5 $ & 3 & 0.43 & $ 0.146 \pm 0.004 $ & 7 & $ 47.88 \pm 0.22 $\\
  & LP~415 &  J04363893+1836567 & M2 $\pm$ 0.5 & 3 & \nodata & $ 3580 \pm 88 $ & $ 0.463 \pm 0.064 $ & $ 650 \pm 70 $ & 21.9 & 8 & 0.40 & $ 0.146 \pm 0.004 $ & 7 & $ 42.109 \pm 0.084 $\\
  & LP5-282 &  J04225989+1318585 & M2 $\pm$ 0.5 & 3 & \nodata & $ 3588 \pm 88 $ & $ 0.491 \pm 0.067 $ & $ 650 \pm 70 $ & $ 9 \pm 2 $ & 3 & 0.16 & $ 0.146 \pm 0.004 $ & 7 & $ 53.29 \pm 0.13 $\\
  & REID~176 &  J04223953+1816097 & M2 $\pm$ 0.5 & 3 & \nodata & $ 3629 \pm 88 $ & $ 0.489 \pm 0.067 $ & $ 650 \pm 70 $ & $ 23.124 \pm 0.009 $\tablenotemark{e} & 9 & 0.44 & $ 0.146 \pm 0.004 $ & 7 & $ 41.707 \pm 0.094 $\\
  & GJ~3290 &  J04271663+1714305 & M2 $\pm$ 0.5 & 3 & \nodata & $ 3663 \pm 88 $ & $ 0.511 \pm 0.070 $ & $ 650 \pm 70 $ & \nodata &  & \nodata & $ 0.146 \pm 0.004 $ & 7 & $ 45.990 \pm 0.098 $\\
Field & GJ~176 & J04425581+1857285 & M2.0 & 10 & \nodata & $ 3591 \pm 98 $ & $ 0.480 \pm 0.060 $ & $2200_{-100}^{+8400}$ & $ 39.3 \pm 0.1 $\tablenotemark{e} & 11 & 0.72 & 0.15 & 12 & $ 9.4730 \pm 0.0063 $\\
  & GJ~436 & J11421096+2642251 & M2.5 & 13 & \nodata & $ 3464 \pm 101 $ & $ 0.424 \pm 0.046 $ & $2400_{-200}^{+4200}$ & $ 44.09 \pm 0.08 $\tablenotemark{e} & 14 & 0.74 & $ -0.0 \pm 0.2 $ & 15 & $ 9.7560 \pm 0.0089 $\\
  & GJ~832 & J21333397-4900323 & M1.5 & 16 & \nodata & $ 3584 \pm 86 $ & $ 0.443 \pm 0.046 $ & $4800_{-2000}^{+5800}$ & $ 46 \pm 9 $ & 11 & 0.59 & $ -0.3 \pm 0.2 $ & 17 & $ 4.9651 \pm 0.0011 $\\
  & G75-55 &  J02582009-0059330 & M1 & 18 & \nodata & $ 3793 \pm 88 $ & $ 0.562 \pm 0.063 $ & $4800_{-3600}^{+5200}$ & \nodata &  & \nodata & $ -0.10 \pm 0.12 $ & 18 & $ 23.896 \pm 0.050 $\\
  & LTT~2050 &  J04374188-1102198 & M2.5 $\pm$ 0.5 & 3 & \nodata & $ 3615 \pm 102 $ & $ 0.459 \pm 0.053 $ & $8600_{-4800}^{+3700}$ & 62 & 19 & 1.2 & $ 0.10 \pm 0.12 $ & 18 & $ 11.2144 \pm 0.0039 $\\
  & GJ~3997 &  J17155010+1900000 & M1 & 18 & \nodata & $ 3713 \pm 74 $ & $ 0.425 \pm 0.033 $ & $8700_{-5400}^{+4000}$ & \nodata &  & \nodata & $ -0.25 \pm 0.12 $ & 18 & $ 13.058 \pm 0.035 $\\
  & GJ~667C & J17185868-3459483 & M1.5 & 20 & \nodata & $ 3489 \pm 58 $ & $ 0.322 \pm 0.020 $ & $8700_{-4400}^{+3400}$ & $ 103.9 \pm 0.7 $\tablenotemark{e} & 11 & 2.1 & $ -0.6 \pm 0.1 $ & 20 & $ 7.2455 \pm 0.0048 $\\
\enddata

\tablenotetext{a}{Alternate spectral type determination from \cite{kraus14} based on optical spectral indices. The types from \cite{kraus14} of the previous column are based on SED fitting.}
\tablenotetext{b}{Stellar effective temperature.}
\tablenotetext{c}{Rotation period.}
\tablenotetext{d}{From \textit{GAIA} \citep{gaia18}.}
\tablenotetext{e}{Period uncertainty excluded from analysis (see Section \ref{sec:obs}).}

\tablerefs{(1) \citealt{kraus14}; (2) \citealt{messina11}; (3) This Work; (4) \citealt{watson06}; (5) \citealt{oelkers18}; (6) \citealt{torres06}; (7) \citealt{cummings17}; (8) \citealt{douglas19}; (9) \citealt{douglas16}; (10) \citealt{forveille09}; (11) \citealt{suarez15}; (12) \citealt{braun14}; (13) \citealt{butler04}; (14) \citealt{bourrier18a}; (15) \citealt{torres08}; (16) \citealt{bailey09}; (17) \citealt{wittenmyer14}; (18) \citealt{gaidos14}; (19) \citealt{astudillo17}; (20) \citealt{anglada13}}

\tablecomments{See Section \ref{sec:obs} for details on stellar parameters estimated in this work as well as the adopted ages.}
\end{deluxetable*}

The stellar sample is limited to subtypes M0-M2.5, with the exception of three stars possibly in the M2.5-M3.5 range, to minimize variations due to differences in stellar properties.
This roughly corresponds to masses of 0.3-0.6~\Msun\ \citep{baraffe96}.
\cite{kraus14} derived spectral types for all but one of the Tuc-Hor objects using both SED fitting and optical spectral indices.
Three Tuc-Hor objects could be K stars according to optical indices, but the inclusion or exclusion of these objects from the analysis has no significant effect (see Section \ref{sec:results}).
For the Hyades and field targets, wherever possible we assigned types ourselves for stars that had spectra from the ELODIE archive at the Observatoire de Haute-Provence (OHP; \citealt{moultaka04})\footnote{\url{http://atlas.obs-hp.fr/elodie/}}, matched against SDSS standards \citep{bochanski07}.
Although the stellar samples from each age group are matched in spectral type, the average mass of the Tuc-Hor group is likely a few 0.01~\Msun\ above that of the Hyades and Field groups because of their pre-main-sequence position on the HR diagram \citep{dantona94}.
We intentionally did not select targets on the basis of existing planets to avoid introducing a bias toward lower activity levels that favor planet detection.
These stars represent a subsample of the stars included in the UV activity evolution analysis of \cite{shkolnik14}, which used data from the \textit{GALEX} survey covering most of the sky.
The stars in the present sample cover a representative range of UV activity levels of M stars at large.

The age of stars in the Tuc-Hor group has been estimated through isochrone fitting, with specific attention paid to the poor performance of stellar structure models at young ages for early types, to be $45\pm4$~Myr \citep{bell15}.
This agrees fairly well with a constraint on the age of 35-45~Myr placed by a lithium depletion argument \citep{kraus14}.
The \Ha\ emission of these stars is also consistent with a young age \citep{kraus14}.
We use an age of $40\pm5$~Myr in this work

The age of the Hyades cluster has been estimated using isochrone and gyrochrone fitting as 700~Myr \citep{maeder81}, $625\pm25$~Myr \citep{perryman98}, and $750\pm100$~Myr \citep{brandt15}.
Recently, \cite{martin18} extended the Li depletion boundary method to brown dwarf masses to derive an age of $650\pm70$~Myr for the Hyades.
For this work, we adopt an age of $650\pm70$~Myr.
The range of possible ages for both the Hyades and Tuc-Hor stars is small relative to the $>2$ orders of magnitude in age spanned by the full stellar sample.

We computed age estimates for the field stars based on combined isochrone and gyrochrone fitting via the \texttt{stardate} package \citep{angus19b}.
For this, we provided \texttt{stardate} with \gaia\ \citep{gaia18} and \textit{2MASS} \citep{cutri03} magnitudes for each field star, along with the rotation periods where available.
The derived ages have uncertainties of several Gyr, as shown in Table \ref{tbl:starprops}.
Such large uncertainties are reasonable given that the rotational evolution of K and M stars is an area of active study, with evidence for complicating variations in the rate of spin-down with age \replaced{\citep{newton16,newton18,curtis19,spada20}}{\citep{newton16,newton18,curtis19,spada20,curtis20}}.

\startlongtable
\begin{deluxetable}{llrrr}
\tablewidth{0pt}
\tabletypesize{\footnotesize}
\tablecaption{Summary of observations.\label{tbl:obssum}}
\tablehead{\colhead{Star} & \colhead{Grating} & \colhead{$T_\mathrm{exp}$\tablenotemark{a}} & \colhead{First} & \colhead{Last}\\
\colhead{} & \colhead{} & \colhead{(ks)} & \colhead{Observation} & \colhead{Observation}}
\startdata
J03315 & G130M & 10.1 & 2017-07-20 & 2017-07-20\\
 & G160M & 10.8 & 2017-07-19 & 2017-07-20\\
 & G230L & 4.6 & 2017-07-19 & 2017-07-19\\
J00240 & G130M & 1.6 & 2017-08-30 & 2017-08-30\\
 & G160M & 1.5 & 2017-08-30 & 2017-08-30\\
 & G230L & 0.3 & 2017-08-30 & 2017-08-30\\
J02543 & G130M & 1.4 & 2017-08-12 & 2017-08-12\\
 & G160M & 1.0 & 2017-08-12 & 2017-08-12\\
 & G230L & 0.4 & 2017-08-12 & 2017-08-12\\
J00393 & G130M & 1.3 & 2017-09-21 & 2017-09-22\\
 & G160M & 1.5 & 2017-09-21 & 2017-09-21\\
 & G230L & 0.6 & 2017-09-21 & 2017-09-21\\
J23261 & G130M & 1.4 & 2017-08-18 & 2017-08-18\\
 & G160M & 1.4 & 2017-08-18 & 2017-08-18\\
 & G230L & 0.6 & 2017-08-18 & 2017-08-18\\
J01521 & G130M & 1.3 & 2017-08-17 & 2017-08-17\\
 & G160M & 1.5 & 2017-08-17 & 2017-08-17\\
 & G230L & 0.3 & 2017-08-17 & 2017-08-17\\
J22025 & G130M & 1.3 & 2017-08-30 & 2017-08-30\\
 & G160M & 1.3 & 2017-08-30 & 2017-08-30\\
 & G230L & 0.3 & 2017-08-30 & 2017-08-30\\
J02125 & G130M & 1.8 & 2017-10-04 & 2017-10-04\\
 & G160M & 0.7 & 2017-10-04 & 2017-10-04\\
 & G230L & 0.3 & 2017-10-04 & 2017-10-04\\
J02365 & G130M & 10.0 & 2017-08-09 & 2017-08-09\\
 & G160M & 10.9 & 2017-08-08 & 2017-08-08\\
 & G230L & 1.8 & 2017-08-08 & 2017-08-08\\
J02001 & G130M & 1.8 & 2017-08-31 & 2017-08-31\\
 & G160M & 0.9 & 2017-08-31 & 2017-08-31\\
 & G230L & 0.4 & 2017-08-31 & 2017-08-31\\
J23285 & G130M & 1.3 & 2017-08-19 & 2017-08-19\\
 & G160M & 1.5 & 2017-08-18 & 2017-08-19\\
 & G230L & 0.3 & 2017-08-18 & 2017-08-18\\
J22463 & G130M & 2.3 & 2017-09-01 & 2017-09-01\\
 & G160M & 1.5 & 2017-09-01 & 2017-09-01\\
 & G230L & 0.6 & 2017-09-01 & 2017-09-01\\
TYC~1265 & G130M & 10.4 & 2018-08-22 & 2018-08-22\\
 & G160M & 10.8 & 2018-08-20 & 2018-08-20\\
 & G230L & 2.2 & 2018-08-20 & 2018-08-20\\
HAN~192 & G130M & 12.2 & 2018-12-02 & 2019-03-30\\
 & G160M & 10.8 & 2018-11-29 & 2018-11-29\\
 & G230L & 2.2 & 2018-11-29 & 2018-11-29\\
LP5-282 & G130M & 7.6 & 2018-02-28 & 2018-02-28\\
 & G160M & 10.8 & 2018-02-27 & 2018-02-27\\
 & G230L & 2.3 & 2018-02-27 & 2018-02-27\\
REID~176\tablenotemark{b} & G230L & 20.8 & 2018-03-20 & 2018-12-15\\
GJ~3290 & G130M & 21.4 & 2017-09-29 & 2017-09-30\\
 & G160M & 21.4 & 2017-09-27 & 2017-09-27\\
 & G230L & 5.4 & 2017-09-26 & 2017-09-26\\
LP~415 & G130M & 21.4 & 2017-09-06 & 2017-09-07\\
 & G160M & 21.4 & 2017-09-05 & 2017-09-06\\
 & G230L & 5.4 & 2017-09-05 & 2017-09-05\\
GJ~176 & G130M & 12.6 & 2015-03-02 & 2015-03-02\\
 & G160M & 5.6 & 2015-02-28 & 2015-02-28\\
 & G230L & 5.8 & 2013-09-07 & 2015-02-28\\
GJ~436 & G130M & 81.1 & 2012-06-23 & 2018-02-28\\
 & G160M & 9.5 & 2012-06-23 & 2015-06-25\\
 & G230L & 4.0 & 2015-06-25 & 2015-06-25\\
GJ~832 & G130M & 15.1 & 2012-07-28 & 2014-10-11\\
 & G160M & 7.7 & 2012-07-28 & 2014-10-10\\
 & G230L & 2.8 & 2014-10-10 & 2014-10-10\\
G75-55 & G130M & 11.6 & 2017-10-16 & 2018-11-04\\
 & G160M & 4.5 & 2017-10-16 & 2018-11-04\\
 & G230L & 0.7 & 2017-10-16 & 2017-10-16\\
LTT~2050 & G130M & 12.6 & 2017-10-17 & 2018-11-03\\
 & G160M & 7.5 & 2017-10-17 & 2018-07-14\\
 & G230L & 0.7 & 2017-10-17 & 2017-10-17\\
GJ~3997 & G130M & 2.3 & 2017-09-18 & 2017-09-18\\
 & G160M & 1.3 & 2017-09-18 & 2017-09-18\\
 & G230L & 0.7 & 2017-09-18 & 2017-09-18\\
GJ~667C & G130M & 4.6 & 2015-08-07 & 2015-08-07\\
 & G160M & 4.5 & 2015-08-07 & 2015-08-07\\
 & G230L & 2.5 & 2015-08-07 & 2015-08-07\\
\enddata

\tablenotetext{a}{Cumulative exposure time.}
\tablenotetext{b}{Not observed with G130M or G160M gratings due to detector flare safety restrictions. }

\tablecomments{Most observations are from program 14784 (PI Shkolnik). Observations of GJ~176, GJ~436, GJ~832, and GJ~667C are from programs 12464 and 13650 (PI France), with additional observations of GJ~436 from programs 14767 (PI Sing) and 15174 (PI Loyd).}
\end{deluxetable}

For consistency and completeness, we estimated the effective temperature (\teff) and radii ($R$) of the sample stars from \gaia\ photometry, expanding upon the procedure from \cite{curtis20}.
We calibrated relationships between \teff, \gaia\ $G$-band (4,000-10,000~\AA) surface flux, and \gaia\ \bprp\ color (4,000-6,500 and 6,500-10,000~\AA) for the well-characterized \md\ sample of \cite{mann15}.
Applying this relationship to the \bprp\ colors of the sample stars yielded \teff\ and $G$-band surface fluxes.
Using the measured $G$-band fluxes, we then derived a photometric radius for the star.
For stars in the 40~Myr Tuc-Hor association, these photometric radii exceed those predicted by MIST stellar models \citep{dotter16,choi16} by 10-60\%.
Stellar evolution models are sensitive to formulations of opacity, convection, and boundary conditions, among other details, yielding inaccurate isochronal ages for young stellar groups \citep{bell15}.
Therefore, we prioritize the photometric radii.

To obtain rotation periods for each star, we located \textit{TESS} photometry wherever possible and conducted a Lomb-Scargle analysis, validating all periods by eye.
We also conducted a literature search for rotation period measurements.
This resulted in rotation period measurements for every star except GJ~3997, G75-55, and GJ~3290.
These stars have been observed by either \textit{TESS} or \textit{K2}, but no rotational modulation was apparent in their light curves.

Where there was overlap, literature periods generally agreed with our analysis of \textit{TESS} data.
However, we encountered a few exceptions.
For J23285, \cite{oelkers18} measured a period of 0.74~d whereas we obtained a rotation period at the second harmonic of $0.37$~d from \textit{TESS} photometry.
We selected the 0.74~d period as more consistent with the star's emission line broadening (with the caveat that narrower lines could be due to a pole-on inclination).
The opposite occurred for J22463, where \cite{oelkers18} measured a period of 0.299~d while we measured a period of 1.65~d, and we selected our measurement as, again, more consistent with the star's emission line broadening as well as the \textit{TESS} lightcurve.
Several discrepant values have been measured for J02365, namely 2.93~d \citep{oelkers18}, 1.54~d \citep{messina10}, 0.74~d \citep{watson06}, and 0.74~d (this work).
We chose 0.74~d since this is the only value for which two measurements agree and this period is clear in the \textit{TESS} lightcurve \added{(Appendix \ref{app:period})}.
Since these discrepant rotation periods are all well within the saturated activity regime (Section~\ref{sec:results}), these choices ultimately have no impact on our analysis.

The rotation period discrepancies highlight the fact that such measurements are always subject to the possibility that the true period is a harmonic of the one identified.
For this reason, high precisions on quoted periods can lead to a false sense of accuracy.
As high precisions do not improve our analysis, we omitted uncertainties on periods \deleted{in Table \ref{tbl:starprops} and} in our analysis where they were below 5\%.
We retain larger uncertainties, such as that of TYC~1265, as they have the potential to significantly increase uncertainty in the fits of Section \ref{subsec:evolfits}.

Rossby number, the ratio of \added{a }star's rotation period to the timescale for convective turnover in its interior, $Ro = P_\mathrm{rot}/\tau_c$, is often used as a proxy for a star's magnetic activity \citep{noyes84}.
As stars age and spin down, their Rossby numbers increase.
There is no method to measure a star's convective turnover timescale.
Instead, \tauc\ must be estimated theoretically (e.g, \citealt{kim96}) or empirically by seeking a form for \tauc\ as a function of stellar mass or color that minimizes scatter in fits to stellar activity decay (e.g., \citealt{noyes84}).
Prescriptions for \tauc\ can vary by factors of a few \citep{brandenburg20}, and without a means of directly measuring \tauc, it is impossible to assess the true accuracy of any such prescription.
Nonetheless, \tauc\ remains useful as a tool to predict the quiescent activity level for a star of arbitrary mass and rotation rate so long as the prescription used for \tauc\ is kept consistent.

We used the prescription of \cite{wright18b} to compute \tauc\ from $V - K_S$ colors for the stars in this work.
The empirical \cite{wright18b} relationship is constrained by roughly 300 objects within the same color range as the present sample, including young, fully convective objects.
We obtained $V$ photometry for the targets from \cite{zacharias12} and $K$ photometry from \cite{cutri03}.
Scatter in the \cite{wright18b} relationship is at the 0.05~dex level.

\section{Data Reduction}
\label{sec:reduction}
For each dataset, we used the method of \cite{loyd18b} to identify flares and anomalies in light curves of the band integrated flux measured by each instrument configuration, excluding regions contaminated by emission from Earth's geocorona.
We then extracted integrated spectra using \texttt{calCOS} v3.3.9 excluding these time ranges to obtain ``quiescence only'' spectra.
An analysis of these flares has already been published for the Tuc-Hor sample \citep{loyd18a} and flares in the remainder of the sample will be the subject of a future work.
These spectra might not represent a true quiescence, as unresolved flares could still be present.
However, we estimated from the flare frequency distribution of \cite{loyd18b} that flares below a typical detection limit for the data would contribute on average $<10$\% ($<0.04$~dex) to the time-integrated flux.

Small offsets in the wavelength solution of a few km~\pers\ are possible between COS exposures \citep{linsky10}.
To mitigate these offsets, we applied kernel density estimation to the wavelength calibrated detector events to identify the peaks of the lines in Table \ref{tbl:lines}.
From these, we determined the offset necessary to shift the wavelength solution of the exposure to the stellar rest frame.
We chose this method versus cross-correlation out of concern that a cross-correlation method would perform poorly for exposures where emission lines represented only a few counts above the background and background noise dominated over most of the spectrum.
Applying these shifts in a second run of \texttt{calCOS} generated ``wavelength-corrected'' spectra.

The \texttt{calCOS} pipeline adds a conservative zero-point error to the fluxed spectrum from each exposure.
When the pipeline coadds dithered exposures, these zero-point errors accumulate to unrealistic levels.
For more accurate errors in coadded spectra, we subtracted (in quadrature) a constant offset in the error such that the final error on the band-integrated spectrum matched that of the Poissonian error, effectively removing the zero-point offset.

The \texttt{calCOS} pipeline will not coadd spectra obtained using differing central wavelength settings.
Differing central wavelengths shift COS spectra by up to a few tens of \AA.
Where necessary, we coadded spectra obtained with the same grating at different central wavelength settings using an in-house code described in Appendix  \ref{app:coadd}.
This was the final step in our data reduction process and yielded spectra comprised of all available data, shifted to the rest frame of the star, and with no measurable contribution from flares.

\section{Analysis}
\label{sec:analysis}

\subsection{Line and Pseudocontinuum Fluxes}
\label{subsec:lines}
The strongest lines in the COS spectra not contaminated by geocoronal emission are \Ciii, \Siiii, \Nv, \Cii, \Siiv, \Civ, \Heii, and \Mgii.
Table~\ref{tbl:lines} provides the wavelengths and peak formation temperatures of these lines.
We fit the components of each of these lines with double Voigt profiles \added{convolved with the COS line spread function (LSF)} and a constant offset for the underlying continuum using the affine-invariant Markov Chain Monte Carlo (MCMC) software \texttt{emcee} \citep{foreman13}.
Single Voigt profiles did not capture broad wings at the base of the lines, even when convolved with the COS \replaced{line spread function (LSF)}{LSF}, whereas including two components yielded a satisfactory fit to both the peak and wings of the lines.
For multiplets, we held the widths of the line components fixed between members of the multiplet but allowed their flux ratios to vary freely.
We fit the \Heii\ multiplet as a single line using average parameters (weighted by oscillator-strength) because these lines are naturally unresolved.

\begin{deluxetable}{lrr}
\tabletypesize{\footnotesize}
\caption{Emission lines fit in analysis. \label{tbl:lines}}
\tablehead{\colhead{Ion} & \colhead{$\lambda_\mathrm{rest}$~\AA} & \colhead{$\log_{10} (T_\mathrm{peak}/\mathrm{K})$\tablenotemark{a}}}
\startdata
\Ciii & 1174.93 & 4.7\\
      & 1175.26 & 4.7\\
      & 1175.59 & 4.7\\
      & 1175.71 & 4.7\\
      & 1175.99 & 4.7\\
      & 1176.37 & 4.7\\
\Siiii & 1206.51 & 4.7\\
\Nv   & 1238.82 & 5.2\\
      & 1242.80 & 5.2\\
\Cii  & 1334.53 & 4.5\\
      & 1335.71 & 4.5\\
\Siiv & 1393.76 & 4.9\\
      & 1402.77 & 4.9\\
\Civ  & 1548.20 & 4.8\\
      & 1550.77 & 4.8\\
\Heii\tablenotemark{b} & 1640.4 & 4.9\\
\Mgii & 2796.35 & 3.6-4.0\\
      & 2803.53 & 3.6-4.0\\
\hline
 & \multicolumn{2}{c}{Integration Bands (\AA)}\\
FUV & \multicolumn{2}{l}{1160-1172, 1250-1290, 1320-1330,}\\
Pseudo- & \multicolumn{2}{l}{1340-1390, 1406-1540,}\\ 
continuum & \multicolumn{2}{l}{1590-1635, 1680-1740}\\
\enddata
\tablenotetext{a}{C line formation temperatures are line center values from \cite{avrett08}, the \Mgii\ range is from \cite{leenaarts13a}, and the remainder are from the \href{http://www.chiantidatabase.org/chianti_linelist.html}{CHIANTI database} \citep{dere09}.}
\tablenotetext{b}{Unresolved multiplet.}

\end{deluxetable}

Low S/N data often yielded implausible line widths far different from the high S/N data.
Therefore, we used the fits from high S/N data to guide priors on line widths that we applied to all fits.
We specified a uniform prior on the combined thermal and turbulent broadening, $\sqrt{2kT/m + v_\mathrm{turb}^2}$, of the fit between 20 and 100~km~\pers\ for the narrow line component and a uniform prior between 100 and 500~km~\pers\ for the broad component.
We appended a Gaussian tail with $\sigma=5$~km~\pers\ to the 20~km~\pers\ end of the boxcar prior for the narrow component and a Gaussian tail with $\sigma=50$~km~\pers\ to the 500~km~\pers\ end  of the prior for the broad component.

Several \Cii\ lines showed absorption from the ISM in the blue component of the doublet, since this is formed by transitions from the ground level.
We fit for this absorption with a single narrow absorption line profile.
The MCMC sampler often jumped to parameters that provided a reasonable fit to most of the data, but missed the ISM absorption.
To avoid this, we specified a uniform probability that the ISM absorption was centered within 30~km~\pers\ of the stellar rest frame, including $\sigma=5$~km~\pers\ Gaussian tails on either end of this range.
We also required that the broadening of the ISM line be less than 10~km~\pers\ based on the ISM observations of \cite{redfield02}.
\deleted{In the fits to emission evolution (Section \ref{subsec:evolfits}), we used only the flux of the redward \Cii\ component since it is not affected by the ISM.}
\explain{We decided to include flux from both components in the final publication because we expect readers using these relationships will want to predict the full C II flux and not just that of one component.}

We observed \Mgii\ emission with the G230L grating, which leaves structure below about 30~km~\pers\ in the lines unresolved.
This includes intrinsic line widths below 30~km~\pers, a self-reversed core, or ISM absorption \citep{redfield02}.
We examined the fits and confirmed that the total flux from the line fits, the key parameter for this study, agrees with a simple integration of the data.
Hence, narrow line widths or self adsorptions not captured by the fits did not impact our results.
The possibility of ISM absorption could impact the intrinsic flux we measured, meriting a more careful treatment that is discussed in Section \ref{subsec:ISM}.

\begin{figure}
\includegraphics{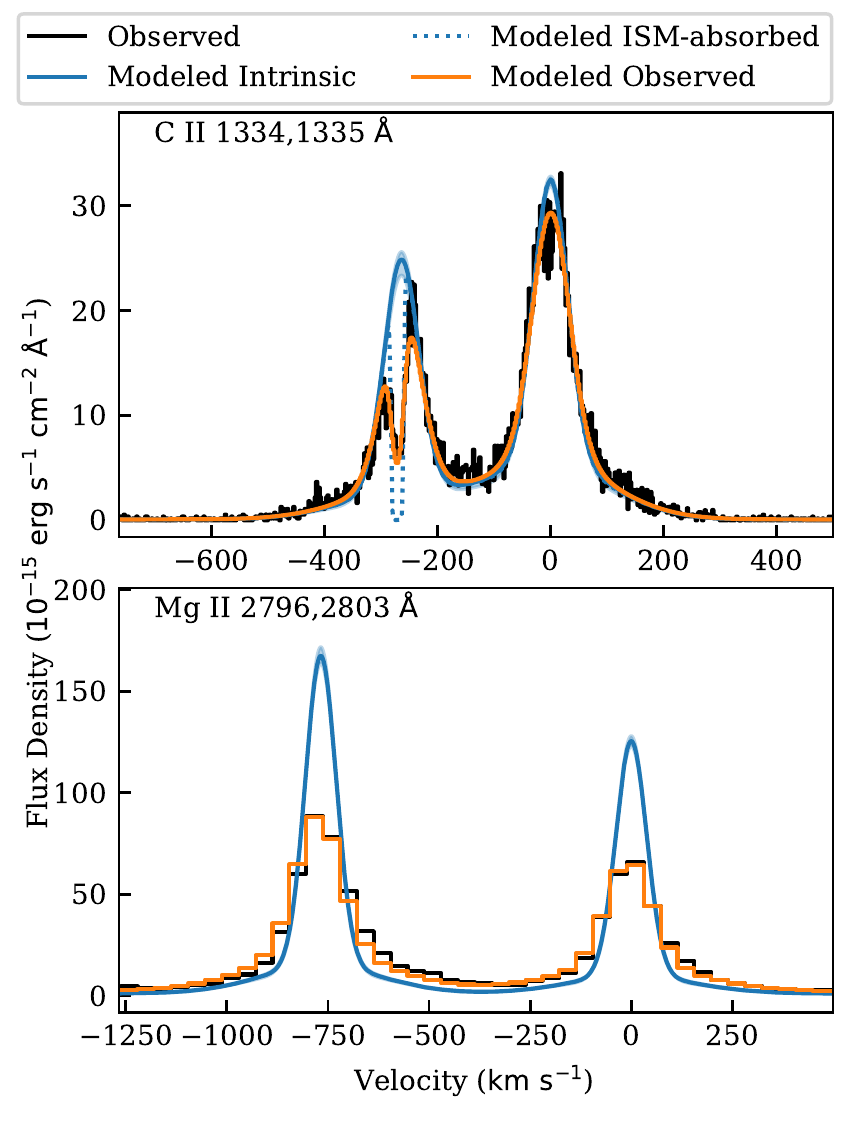}
\caption{Example fits to two UV lines for the star J02365 using the superposition of two Voigt profiles for each component of the doublet.
Top: \Cii\ lines exhibiting clear ISM absorption in the blueward line.
These are well-resolved by the COS~G130M grating ($R \approx 15,000$).
Bottom:
\Mgii\ lines.
The COS~G230L grating ($R \approx 3,000$) used for these observations significantly broadens the intrinsic line profiles and weakens their peaks.
ISM absorption affects \Mgii\ (see Section~\ref{subsec:ISM}), but we did not model any absorption since it was not resolved.
The \Mgii\ lines are the only strong UV lines included in the analysis that were not resolved.
\deleted{When analyzing the evolution of line emission, we excluded the blueward \Cii\ line components since these could be significantly influenced by ISM absorption.}
\label{fig:example}}
\end{figure}

Figure~\ref{fig:example} plots example \Cii\ and \Mgii\ line fits from the sample.
Parameters of the fits and MCMC chains for each line of each star are available from the corresponding author upon request.

In addition to line fluxes, we computed the flux from FUV pseudocontinuum regions between these lines through simple integration.
We use the term pseudocontinuum to deliberately acknowledge that these regions consist primarily of unresolved lines and ionization edges rather than blackbody emission \citep{peacock19b}.
The wavelengths across which we integrated the FUV pseudocontinuum are listed in Table \ref{tbl:lines}.

\subsection{Evolution of Emission Lines}
\label{subsec:evolfits}
We fit an empirical model to the quiescent activity evolution of the stars as a function of age, rotation period, and Rossby number\added{, again using \texttt{emcee} \citep{foreman13}}.
The form of the model was a constant saturation followed by a power law decline,
\fiteqn
where $x$ stands in for age, rotation period, or Rossby number and the free parameters are the saturation surface flux, $F_\mathrm{sat}$; end of saturation, $x_\mathrm{sat}$; and power-law index, $\alpha$.
We included a hyperparameter, $\sigma$, to describe a uniform Gaussian scatter about the fit in log space.

The three age groups of the data do not resolve the transition from saturation to declining activity\replaced{, but this transition has been constrained with previous UV and X-ray data from \textit{GALEX} and \textit{ROSAT} \citep{shkolnik14}.}{.
This lifetime has been probed by a number of previous studies using \Ha, UV, and X-ray observations \citep{west08,jackson12,shkolnik14,johnstone20b}.}
\deleted{Recently, }\cite{mcdonald19} fit \replaced{the}{\textit{ROSAT}} X-ray data \replaced{from}{compiled by} \cite{shkolnik14} \replaced{with the same model}{using MCMC methods}, yielding a posterior distribution on the duration of the saturated period that ranges from 100 to 500~Myr at the \nth{16} and \nth{84} percentiles and favors 400~Myr.
Because age was sparsely sampled by the HAZMAT \textit{HST} program, we applied this posterior as a prior on saturation lifetime in our MCMC fits.
However, doing so yielded a statistically insignificant change in the best-fit value of the saturation lifetime relative to an uninformative prior, indicating that even with sparse age sampling the HAZMAT data constrain the activity saturation lifetime for early M stars.
In fitting quiescent activity vs. rotation period or Rossby number, rather than age, we applied no prior on saturation lifetime.

We included uncertainties in age, rotation period (when significant), and Rossby number in this analysis by computing the data likelihood as an integral of the probability over the path of the model.
Systematic errors in the determination of the convective turnover time could lead to additional uncertainty in Rossby number that we did not include (see Section~\ref{sec:obs}).


\section{Results}
\label{sec:results}
\begin{figure*}
\includegraphics{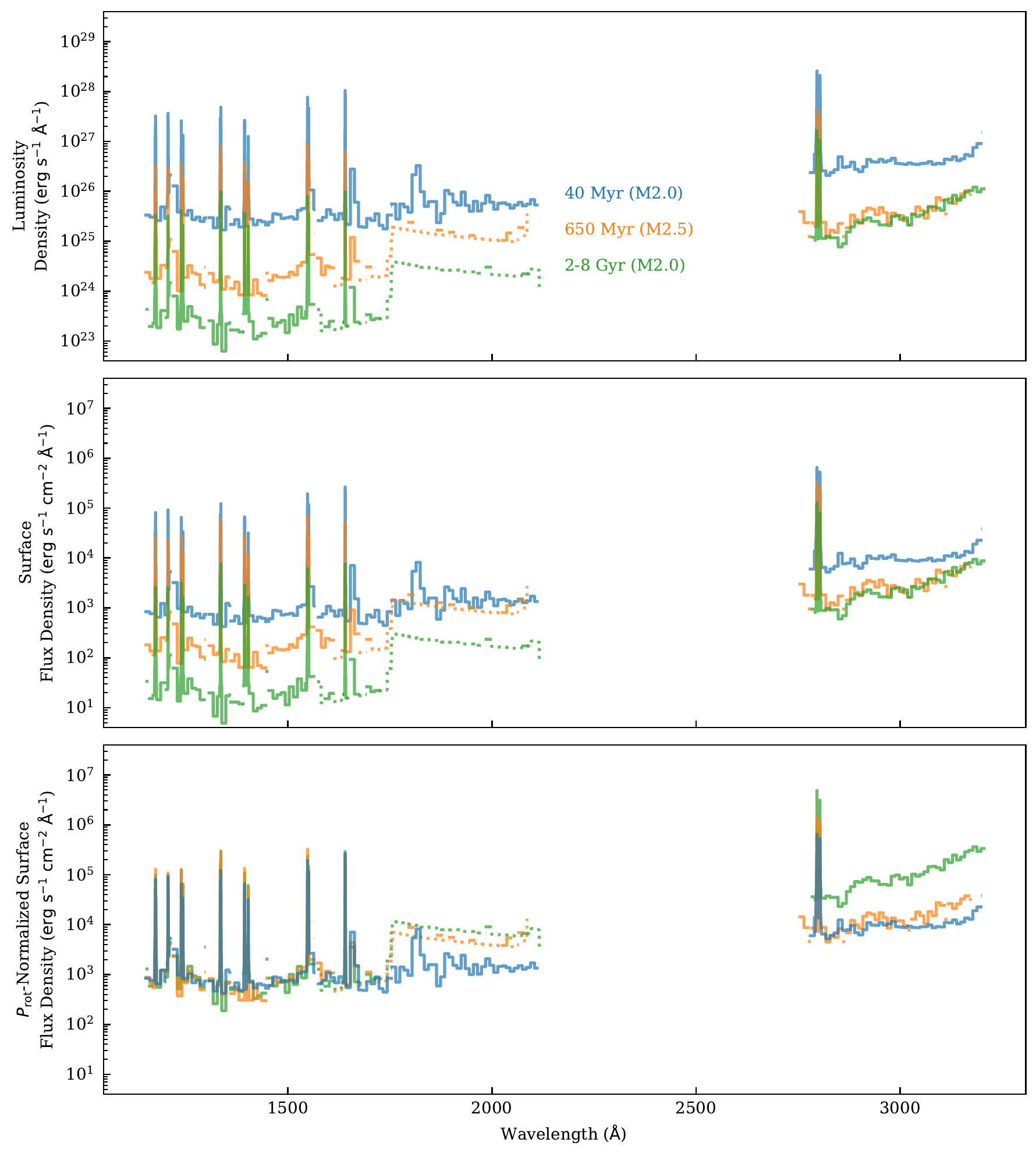}
\caption{Spectra of representative stars from the stellar sample.
The spectra of these fall roughly in the middle of the range of surface fluxes for each group.
Strong emission lines have been replaced with their fits and the regions between binned to 10~\AA\ for higher S/N (see text for more details).
Dotted portions of the spectra represent 2$\sigma$ upper limits where S/N was $< 2$.
The evolution of the UV luminosities (top panel) and surface fluxes (middle panel) with age over roughly two orders of magnitude is clear.
Normalizing by rotation period (bottom panel) results in close consistency between the spectra blueward of 1700~\AA.
See the text for further discussion, including of the disparities in the NUV continuum redward of 2700~\AA.
\label{fig:spectra}}
\end{figure*}

Spectra of representative stars from each age group are shown in Figure~\ref{fig:spectra}, cast in luminosity, surface flux, and a ``rotation-normalized'' surface flux, further explained in Section~\ref{sec:discussion}.
For visual comparison, we split these spectra into continuum regions and strong emission lines and excluded regions contaminated by Earth's geocoronal emission (namely, the \lya\ and \Oi\ lines).
We replaced the emission lines with their double-Voigt profiles from MCMC fits with instrumental broadening removed.
We also removed ISM absorption from the blueward \Cii\ line.
For the regions between these lines, we binned to 10~\AA\ for higher S/N.
The spectra consist of data from the COS G130M and G160M gratings ($R$ = 12,000-20,000) up to roughly 1750~\AA, beyond which the data originate from the G230L grating ($R$ = 2,000-4,000).
For many spectra, this results in a jump to higher noise levels redward of 1750~\AA, yielding a nondetection of the continuum represented in the figure as dotted lines giving 2$\sigma$ upper limits.
The G230L mode does not capture flux from roughly 2200 to 2750~\AA, hence the gap in the spectra.

\begin{deluxetable*}{lrrrrrrrrr}
\tablewidth{0pt}
\tabletypesize{\footnotesize}
\rotate
\tablecaption{UV Emission line surface fluxes in $10^4$ $\mathrm{erg\ s^{-1}\ cm^{-2}}$.\label{tbl:fluxes}}
\tablehead{\colhead{Star} & \colhead{Mg II} & \colhead{C II} & \colhead{Si III } & \colhead{C III} & \colhead{Si IV} & \colhead{He II} & \colhead{C IV} & \colhead{N V} & \colhead{Psuedocont.\tablenotemark{a}}}
\startdata
J03315 & $83.6_{-0.8}^{+2.8}$ & $9.1_{-1.0}^{+0.9}$ & $3.68_{-0.09}^{+0.17}$ & $7.78_{-0.25}^{+0.47}$ & $4.10_{-0.12}^{+0.10}$ & $13.31_{-0.34}^{+0.19}$ & $16.59_{-0.18}^{+0.39}$ & $4.9_{-0.4}^{+1.0}$ & $ 26.74 \pm 0.23 $\\
J00240 & $87.3_{-2.3}^{+1.9}$ & $12.60_{-0.25}^{+0.18}$ & $5.93_{-0.09}^{+0.11}$ & $9.45_{-0.23}^{+0.18}$ & $6.53_{-0.23}^{+0.18}$ & $12.86_{-0.21}^{+0.27}$ & $18.36_{-0.33}^{+0.30}$ & $4.130_{-0.087}^{+0.083}$ & $ 26.15 \pm 0.73 $\\
J02543 & $51.0_{-2.2}^{+1.9}$ & $5.39_{-0.25}^{+0.11}$ & $2.39_{-0.15}^{+0.44}$ & $4.73_{-0.25}^{+0.31}$ & $3.53_{-0.31}^{+0.22}$ & $7.16_{-0.17}^{+0.22}$ & $12.3_{-2.0}^{+0.6}$ & $3.00_{-0.44}^{+0.34}$ & $ 15.95 \pm 0.71 $\\
J23261 & $167.7_{-3.0}^{+3.1}$ & $11.67_{-0.39}^{+0.23}$ & $4.50_{-0.13}^{+0.16}$ & $8.39_{-0.36}^{+0.43}$ & $4.82_{-0.15}^{+0.17}$ & $15.94_{-0.23}^{+0.29}$ & $20.51_{-0.17}^{+0.20}$ & $3.63_{-0.12}^{+0.14}$ & $ 29.3 \pm 1.0 $\\
J01521 & $34.3_{-1.7}^{+1.5}$ & $6.0_{-1.5}^{+0.5}$ & $2.12_{-0.19}^{+0.46}$ & $6.8_{-0.4}^{+1.3}$ & $3.17_{-0.17}^{+0.56}$ & $9.93_{-0.52}^{+0.80}$ & $9.89_{-0.28}^{+0.29}$ & $2.40_{-0.18}^{+0.26}$ & $ 15.17 \pm 0.79 $\\
J00393 & $62.8_{-1.5}^{+1.6}$ & $7.25_{-0.22}^{+0.57}$ & $3.71_{-0.82}^{+0.20}$ & $6.84_{-0.29}^{+0.65}$ & $4.06_{-0.15}^{+0.13}$ & $10.13_{-0.13}^{+0.29}$ & $16.34_{-0.24}^{+0.57}$ & $3.96_{-0.27}^{+0.16}$ & $ 21.21 \pm 0.80 $\\
J02001 & $ 77.2 \pm 1.6 $ & $9.69_{-0.31}^{+0.28}$ & $3.85_{-0.14}^{+0.26}$ & $7.71_{-0.32}^{+0.49}$ & $4.38_{-0.21}^{+0.87}$ & $15.56_{-0.28}^{+0.29}$ & $26.6_{-0.6}^{+1.1}$ & $4.56_{-0.22}^{+0.34}$ & $ 33.03 \pm 0.98 $\\
J02365 & $146.8_{-2.4}^{+1.6}$ & $11.05_{-0.11}^{+0.24}$ & $4.182_{-0.039}^{+0.037}$ & $6.49_{-0.17}^{+0.24}$ & $5.32_{-0.24}^{+0.12}$ & $13.94_{-0.16}^{+0.15}$ & $16.85_{-0.27}^{+0.32}$ & $3.889_{-0.098}^{+0.089}$ & $ 25.88 \pm 0.20 $\\
J22025\tablenotemark{b} & $107.2_{-4.8}^{+4.0}$ & $ 16.48 \pm 0.24 $ & $10.47_{-0.22}^{+0.19}$ & $17.33_{-0.45}^{+0.41}$ & $14.70_{-0.25}^{+0.26}$ & $11.30_{-0.39}^{+0.23}$ & $17.44_{-0.26}^{+0.13}$ & $ 5.89 \pm 0.16 $ & $ 28.63 \pm 0.98 $\\
J23285 & $150.8_{-3.2}^{+4.9}$ & $11.16_{-0.27}^{+0.55}$ & $ 3.87 \pm 0.13 $ & $9.13_{-0.89}^{+0.48}$ & $5.3_{-0.3}^{+2.0}$ & $17.33_{-0.20}^{+0.25}$ & $16.75_{-0.23}^{+0.20}$ & $4.61_{-0.24}^{+0.20}$ & $ 31.1 \pm 1.1 $\\
J22463 & $77_{-14}^{+13}$ & $10.24_{-0.42}^{+0.37}$ & $3.09_{-0.18}^{+0.78}$ & $5.95_{-0.42}^{+0.69}$ & $4.49_{-0.31}^{+0.22}$ & $9.67_{-0.17}^{+0.12}$ & $12.7_{-0.5}^{+2.2}$ & $3.92_{-0.30}^{+0.41}$ & $ 21.4 \pm 1.1 $\\
J02125 & \nodata & $8.66_{-0.14}^{+0.17}$ & $3.83_{-0.19}^{+0.09}$ & $5.37_{-0.15}^{+0.23}$ & $4.97_{-0.89}^{+0.34}$ & $18.0_{-6.4}^{+3.6}$ & $15.70_{-0.24}^{+0.28}$ & $3.14_{-0.18}^{+0.67}$ & $ 24.2 \pm 1.3 $\\
HAN~192 & $ 58.3 \pm 1.3 $ & $1.786_{-0.056}^{+0.044}$ & $0.556_{-0.025}^{+0.052}$ & $1.12_{-0.17}^{+0.09}$ & $0.76_{-0.12}^{+0.16}$ & $1.77_{-0.08}^{+0.11}$ & $2.91_{-0.09}^{+0.10}$ & $0.87_{-0.11}^{+0.04}$ & $ 3.87 \pm 0.28 $\\
TYC~1265 & $59.3_{-1.9}^{+1.8}$ & $1.900_{-0.052}^{+0.044}$ & $0.62_{-0.10}^{+0.04}$ & $1.38_{-0.09}^{+0.12}$ & $0.82_{-0.14}^{+0.10}$ & $2.66_{-0.44}^{+0.95}$ & $2.94_{-0.10}^{+0.11}$ & $0.84_{-0.08}^{+0.18}$ & $ 5.74 \pm 0.34 $\\
LP~415 & $39.90_{-0.52}^{+0.68}$ & $1.857_{-0.030}^{+0.074}$ & $0.65_{-0.07}^{+0.10}$ & $ 1.60 \pm 0.12 $ & $0.85_{-0.09}^{+0.12}$ & $1.740_{-0.054}^{+0.068}$ & $4.57_{-0.19}^{+0.28}$ & $0.94_{-0.08}^{+0.18}$ & $ 4.38 \pm 0.23 $\\
LP5-282 & $119.4_{-2.0}^{+2.3}$ & $11.96_{-0.27}^{+0.19}$ & $4.29_{-0.15}^{+0.09}$ & $6.8_{-0.5}^{+1.1}$ & $3.6_{-0.4}^{+1.4}$ & $8.98_{-0.17}^{+0.15}$ & $16.4_{-1.4}^{+1.2}$ & $5.66_{-0.30}^{+0.43}$ & $ 25.48 \pm 0.63 $\\
REID~176\tablenotemark{c} & $26.09_{-0.29}^{+0.27}$ & \nodata & \nodata & \nodata & \nodata & \nodata & \nodata & \nodata & \nodata\\
GJ~3290 & $40.61_{-0.78}^{+0.74}$ & $1.276_{-0.020}^{+0.021}$ & $0.364_{-0.013}^{+0.017}$ & $ 0.864 \pm 0.044 $ & $0.416_{-0.021}^{+0.029}$ & $1.20_{-0.07}^{+0.23}$ & $2.20_{-0.06}^{+0.50}$ & $0.66_{-0.05}^{+0.12}$ & $ 3.99 \pm 0.22 $\\
GJ~176 & $17.63_{-0.09}^{+0.11}$ & $0.501_{-0.008}^{+0.010}$ & $0.1395_{-0.0067}^{+0.0034}$ & $ 0.1925 \pm 0.0061 $ & $0.1592_{-0.0031}^{+0.0028}$ & $0.5332_{-0.0062}^{+0.0040}$ & $0.857_{-0.024}^{+0.016}$ & $0.2522_{-0.0053}^{+0.0047}$ & $ 1.300 \pm 0.039 $\\
GJ~436 & $6.05_{-0.13}^{+0.11}$ & $0.1231_{-0.0045}^{+0.0024}$ & $0.05045_{-0.00067}^{+0.00069}$ & $0.0572_{-0.0010}^{+0.0011}$ & $0.0690_{-0.0018}^{+0.0019}$ & $0.1401_{-0.0024}^{+0.0031}$ & $0.2566_{-0.0057}^{+0.0056}$ & $0.0999_{-0.0024}^{+0.0015}$ & $ 0.728 \pm 0.029 $\\
GJ~832 & $6.506_{-0.047}^{+0.036}$ & $0.1311_{-0.0012}^{+0.0010}$ & $0.0420_{-0.0011}^{+0.0015}$ & $0.0471_{-0.0008}^{+0.0011}$ & $0.05333_{-0.00084}^{+0.00088}$ & $0.1828_{-0.0015}^{+0.0020}$ & $0.2129_{-0.0031}^{+0.0036}$ & $ 0.0851 \pm 0.0010 $ & $ 0.494 \pm 0.011 $\\
G75-55 & $ 33.6 \pm 1.4 $ & $0.576_{-0.018}^{+0.013}$ & $0.233_{-0.010}^{+0.047}$ & $0.429_{-0.028}^{+0.039}$ & $0.271_{-0.007}^{+0.019}$ & $0.892_{-0.022}^{+0.046}$ & $1.04_{-0.03}^{+0.21}$ & $0.32_{-0.06}^{+0.11}$ & $ 1.85 \pm 0.12 $\\
LTT~2050 & $ 7.91 \pm 0.32 $ & $0.1839_{-0.0043}^{+0.0027}$ & $0.0785_{-0.0078}^{+0.0048}$ & $0.1200_{-0.0092}^{+0.0095}$ & $0.0813_{-0.0082}^{+0.0067}$ & $0.33_{-0.10}^{+0.03}$ & $0.365_{-0.009}^{+0.020}$ & $0.134_{-0.017}^{+0.010}$ & $ 0.586 \pm 0.032 $\\
GJ~3997 & $6.02_{-0.43}^{+0.84}$ & $0.1057_{-0.0072}^{+0.0071}$ & $0.035_{-0.016}^{+0.021}$ & $ 0.233 \pm 0.017 $ & $0.080_{-0.008}^{+0.062}$ & $0.268_{-0.018}^{+0.035}$ & $0.167_{-0.016}^{+0.019}$ & $0.052_{-0.007}^{+0.012}$ & $ 2.60 \pm 0.11 $\\
GJ~667C & $ 6.02 \pm 0.12 $ & $0.1098_{-0.0034}^{+0.0033}$ & $0.0619_{-0.0056}^{+0.0053}$ & $0.0794_{-0.0023}^{+0.0029}$ & $0.0892_{-0.0026}^{+0.0028}$ & $0.2139_{-0.0034}^{+0.0043}$ & $0.3344_{-0.0050}^{+0.0071}$ & $0.0762_{-0.0020}^{+0.0019}$ & $ 0.997 \pm 0.044 $\\
\enddata

\tablenotetext{a}{FUV Pseudocontinuum flux integrated over the ranges given in Table \ref{tbl:lines}.}
\tablenotetext{b}{Line fluxes affected by a double-peaked morphology not captured by the fits. Possible binary. Excluded from evolution fits.}
\tablenotetext{c}{FUV lines not observed due to detector flare safety regulations.}

\end{deluxetable*}

\begin{figure*}
\includegraphics{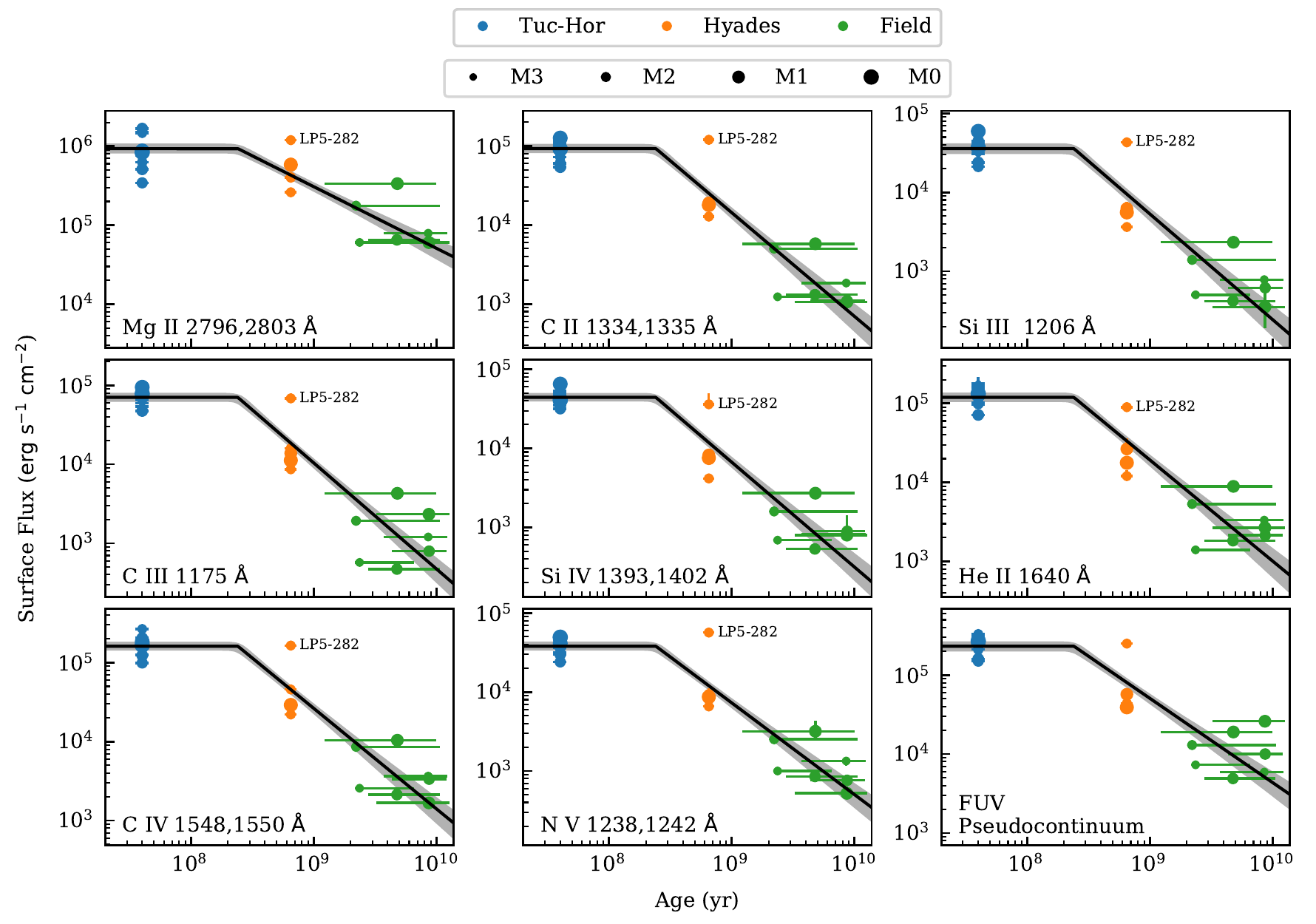}
\caption{Surface fluxes of strong UV emission lines as a function of age.
\evolCaption\
LP5-282 is highlighted as an outlier in these relationships, but is no longer an outlier in the rotation and Rossby number relationships (Figures \ref{fig:PEvol} and \ref{fig:RoEvol}).
\label{fig:ageEvol}}
\end{figure*}

\begin{figure*}
\includegraphics{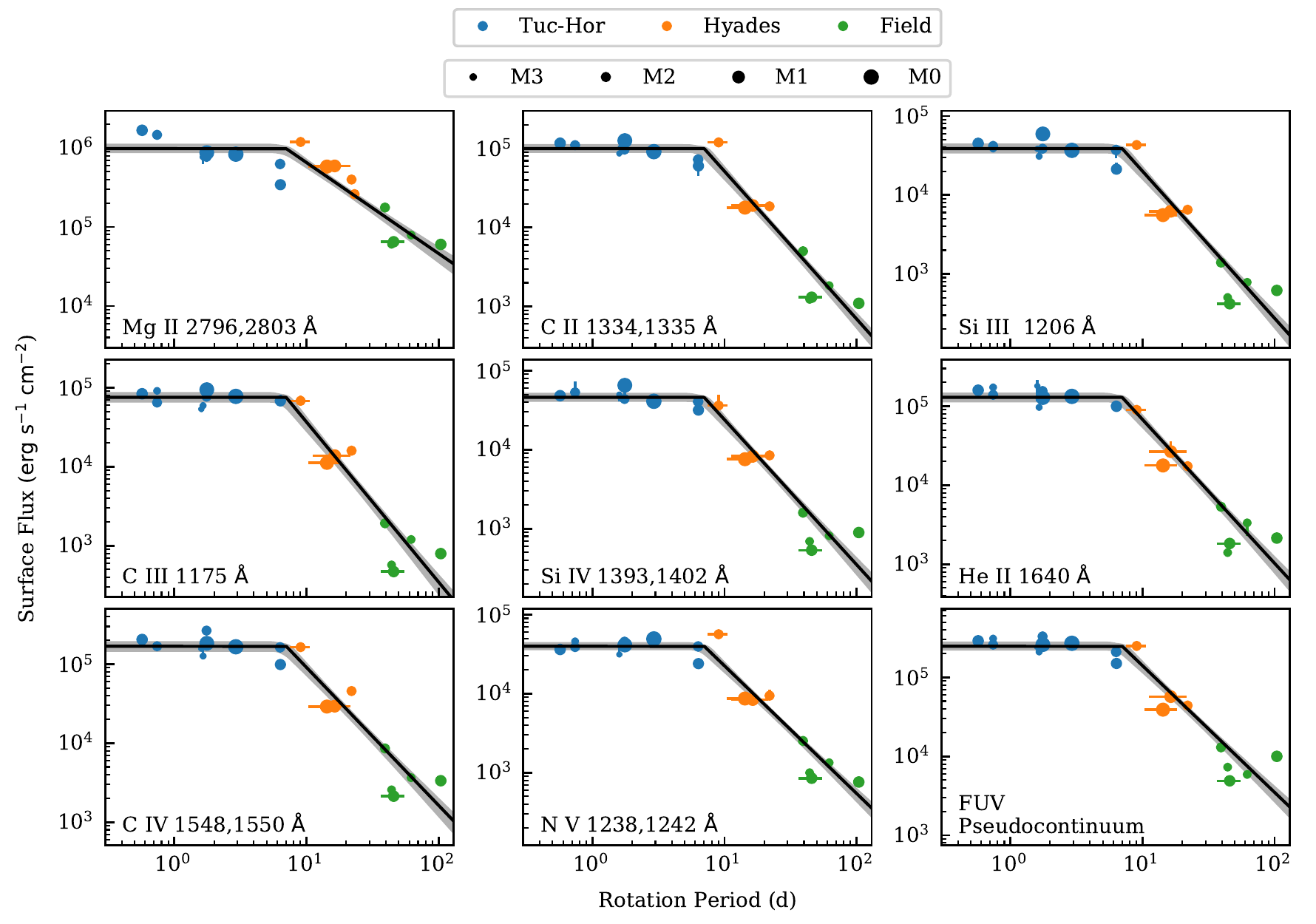}
\caption{Surface fluxes of strong UV emission lines as a function of rotation period.
\evolCaption
\label{fig:PEvol}}
\end{figure*}

\begin{figure*}
\includegraphics{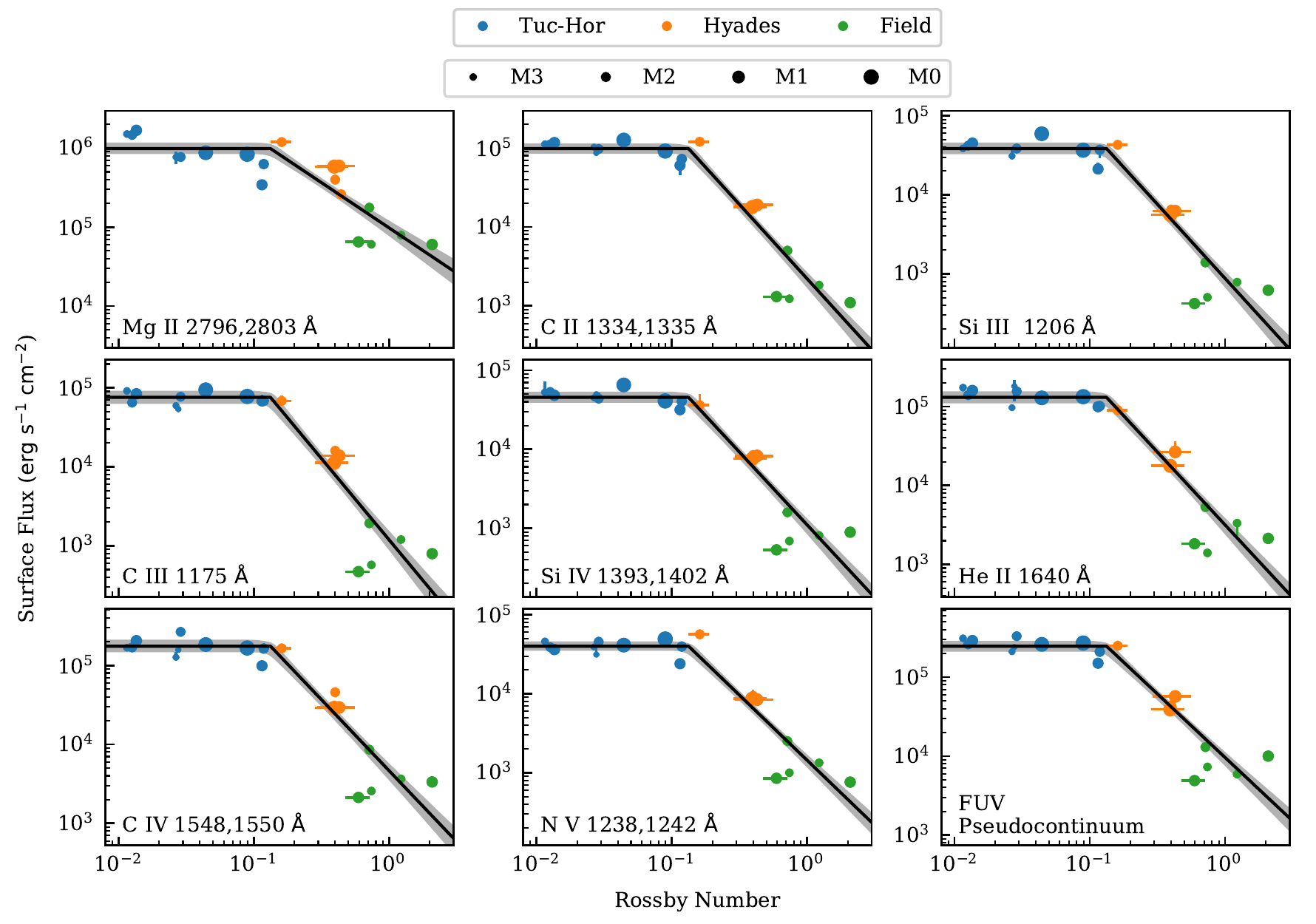}
\caption{Surface fluxes of strong UV emission lines as a function of Rossby number.
\evolCaption
\label{fig:RoEvol}}
\end{figure*}

Figures \ref{fig:ageEvol}, \ref{fig:PEvol}, and \ref{fig:RoEvol} show the results of our fits to the age, rotation period, and Rossby number evolution of UV emission line fluxes.
The emission line fluxes driving these fits, computed from our double-Voigt profiles, are given in Table~\ref{tbl:fluxes}.
Table \ref{tbl:evol} gives the parameters of the fits to emission evolution.
In addition to the fits shown, we computed fits excluding the possible K stars in the Tuc-Hor sample.
The changes this produced were below the $1\sigma$ uncertainties in the fit parameters, generally increasing power law indices by a few hundredths to a tenth and slightly increasing the $\sigma$  hyperparameter.

\begin{deluxetable}{lrrrrrr}
\tablewidth{0pt}
\tabletypesize{\footnotesize}
\tablecaption{Fits of emission line surface flux evolution with age, rotation, and Rossby number.\label{tbl:evol}}
\tablehead{
\colhead{Line} &
\colhead{$\log_{10}(F_\mathrm{sat})$} &
\colhead{$\alpha$} &
\colhead{$\sigma$} &
\colhead{$\hat{\sigma}_\mathrm{TH}$} &
\colhead{$\hat{\sigma}_\mathrm{Hy}$} &
\colhead{$\hat{\sigma}_\mathrm{Fld}$}
\\
 &
\colhead{$\mathrm{erg\ s^{-1}\ cm^{-2}}$} &
 &
 }
\startdata
\hline
\multicolumn{7}{c}{Age}\\
\multicolumn{7}{c}{$\log_{10}(x_\mathrm{sat}/\mathrm{yr}) = 8.381_{-0.050}^{+0.046}$}\\
\hline
Mg II & $5.970_{-0.063}^{+0.064}$ & $0.78_{-0.09}^{+0.11}$ & 0.21 & 0.21 & 0.20 & 0.28\\
C II & $4.967_{-0.059}^{+0.060}$ & $1.31_{-0.12}^{+0.14}$ & 0.21 & 0.11 & 0.35 & 0.32\\
Si III  & $4.554_{-0.071}^{+0.068}$ & $1.34_{-0.14}^{+0.15}$ & 0.23 & 0.12 & 0.37 & 0.35\\
C III & $4.847_{-0.056}^{+0.058}$ & $1.34_{-0.12}^{+0.14}$ & 0.19 & 0.09 & 0.31 & 0.48\\
Si IV & $4.646_{-0.053}^{+0.057}$ & $1.33_{-0.13}^{+0.14}$ & 0.19 & 0.08 & 0.31 & 0.35\\
He II & $5.077_{-0.060}^{+0.061}$ & $1.28_{-0.13}^{+0.16}$ & 0.21 & 0.12 & 0.30 & 0.39\\
C IV & $5.208_{-0.057}^{+0.058}$ & $1.27_{-0.12}^{+0.14}$ & 0.20 & 0.11 & 0.31 & 0.33\\
N V & $4.580_{-0.057}^{+0.060}$ & $1.16_{-0.11}^{+0.13}$ & 0.19 & 0.09 & 0.34 & 0.27\\
Cont.\tablenotemark{a} & $5.365_{-0.067}^{+0.061}$ & $1.07_{-0.11}^{+0.14}$ & 0.21 & 0.11 & 0.31 & 0.38\\
\hline
\multicolumn{7}{c}{Rotation Period}\\
\multicolumn{7}{c}{$\log_{10}(x_\mathrm{sat}/\mathrm{d}) = 0.846_{-0.085}^{+0.040}$}\\
\hline
Mg II & $5.993_{-0.058}^{+0.062}$ & $1.15_{-0.12}^{+0.13}$ & 0.19 & 0.20 & 0.07 & 0.18\\
C II & $5.000_{-0.055}^{+0.056}$ & $1.87_{-0.14}^{+0.15}$ & 0.17 & 0.09 & 0.18 & 0.26\\
Si III  & $4.588_{-0.062}^{+0.064}$ & $ 1.86 \pm 0.16 $ & 0.19 & 0.11 & 0.20 & 0.31\\
C III & $4.878_{-0.068}^{+0.069}$ & $2.03_{-0.16}^{+0.17}$ & 0.22 & 0.08 & 0.20 & 0.36\\
Si IV & $4.662_{-0.058}^{+0.057}$ & $1.84_{-0.16}^{+0.15}$ & 0.18 & 0.08 & 0.15 & 0.31\\
He II & $5.112_{-0.063}^{+0.060}$ & $1.82_{-0.16}^{+0.15}$ & 0.19 & 0.10 & 0.14 & 0.32\\
C IV & $5.227_{-0.071}^{+0.065}$ & $1.75_{-0.15}^{+0.16}$ & 0.21 & 0.11 & 0.21 & 0.30\\
N V & $4.601_{-0.050}^{+0.052}$ & $ 1.62 \pm 0.13 $ & 0.15 & 0.09 & 0.20 & 0.21\\
Cont.\tablenotemark{a} & $5.395_{-0.056}^{+0.059}$ & $1.60_{-0.14}^{+0.15}$ & 0.18 & 0.09 & 0.18 & 0.30\\
\hline
\multicolumn{7}{c}{Rossby Number}\\
\multicolumn{7}{c}{$\log_{10}(x_\mathrm{sat}) = -0.876_{-0.061}^{+0.037}$}\\
\hline
Mg II & $5.994_{-0.070}^{+0.077}$ & $ 1.15 \pm 0.16 $ & 0.23 & 0.20 & 0.12 & 0.24\\
C II & $4.992_{-0.068}^{+0.065}$ & $ 1.88 \pm 0.16 $ & 0.22 & 0.09 & 0.04 & 0.37\\
Si III  & $4.586_{-0.073}^{+0.077}$ & $1.88_{-0.17}^{+0.18}$ & 0.23 & 0.11 & 0.06 & 0.42\\
C III & $4.879_{-0.083}^{+0.085}$ & $ 2.06 \pm 0.20 $ & 0.27 & 0.08 & 0.09 & 0.49\\
Si IV & $4.657_{-0.067}^{+0.073}$ & $1.84_{-0.15}^{+0.17}$ & 0.22 & 0.08 & 0.05 & 0.42\\
He II & $ 5.116 \pm 0.075 $ & $1.84_{-0.15}^{+0.16}$ & 0.23 & 0.10 & 0.11 & 0.42\\
C IV & $5.243_{-0.077}^{+0.081}$ & $ 1.80 \pm 0.17 $ & 0.25 & 0.11 & 0.08 & 0.42\\
N V & $4.603_{-0.059}^{+0.062}$ & $ 1.65 \pm 0.14 $ & 0.19 & 0.09 & 0.06 & 0.32\\
Cont.\tablenotemark{a} & $5.392_{-0.070}^{+0.068}$ & $1.61_{-0.15}^{+0.16}$ & 0.22 & 0.09 & 0.09 & 0.39\\
\enddata

\tablenotetext{a}{FUV pseudocontinuum.}
\tablecomments{Fits follow the form \fiteqn where $F$ is surface flux, $x$ is a placeholder for age [yr], rotation period [d], or Rossby number [dimensionless], and $\mathrm{sat}$ stands for saturation. All lines share the same $x_\mathrm{sat}$ and were fit simultaneously. The intrinsic scatter, $\sigma$, was included as a hyperparameter to the fit. The sample standard deviation of points about the median fit, $\hat{\sigma}$, is also included for the Tuc-Hor (TH), Hyades (Hy), and Field (Fld) groups separately.}
\end{deluxetable}


\section{Discussion}
\label{sec:discussion}

The UV emission of early M stars shows a clear evolution with age and rotation (Figure~\ref{fig:spectra}).
This is true of both strong emission lines as well as the regions between lines, a pseudocontinuum of weak or unresolved lines and atomic and molecular bound-free opacities \citep{peacock19b}.
For most lines, emission declines by around two orders of magnitude from youth, when emission levels are saturated, to several Gyr.
The spectrally resolved evolution we measured refines and extends previously established relationships between broadband X-ray and UV emission and age \replaced{\citep{jackson12,shkolnik14,mcdonald19}}{\citep{feigelson04,preibisch05,jackson12,stelzer13,stelzer14,shkolnik14,mcdonald19}}.
These spectra will enable improved modeling of the inobservable EUV emission of stars, such as \cite{peacock20} carried out using broadband UV measurements.

\subsection{Where Emission Forms Influences How It Evolves}

\begin{figure}
\includegraphics{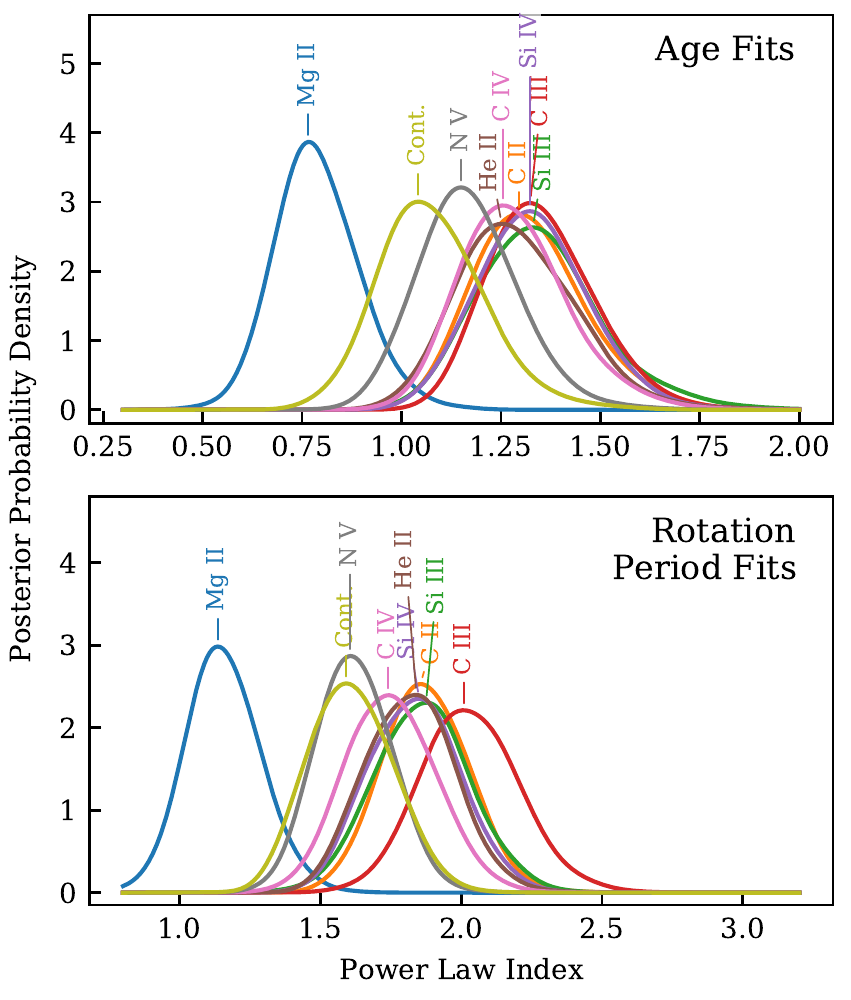}
\caption{Smoothed posterior distributions for the index of the power law decline in quiescent activity for strong UV emission lines.
Most are statistically indistinguishable.
However, \Mgii\ shows a significantly less rapid decline (lower index) relative to \Ciii\ and most other FUV lines, likely because a fraction of \Mgii\ emission originates from radiative excitement of ions in the stellar photosphere rather than the activity-heated upper atmosphere.
\label{fig:indices}}
\end{figure}

\begin{figure}
\includegraphics{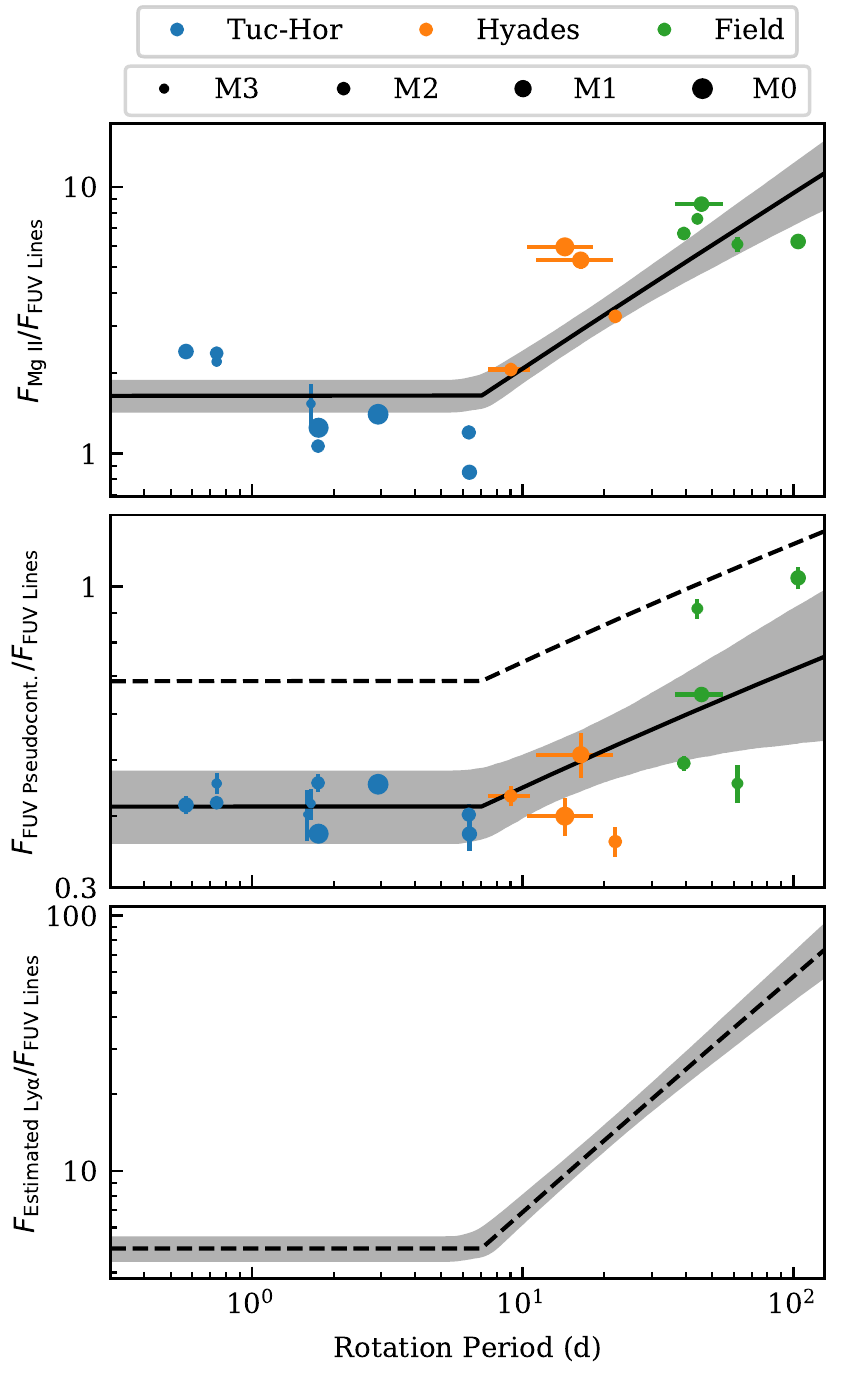}
\caption{Evolution of UV flux ratios as early M stars spin down.
The slower evolution of \Mgii\ and pseudocontinuum flux corresponds to an increase in these ratios with rotation.
``FUV lines'' refers to every strong line (i.e., commensurate with \Siiv) in the 1160-1740~\AA\ range except \Oi\deleted{, the blue component of \Cii,} and \Alii.
Including these would increase the summed FUV line flux by $\sim$10\%.
\Mgii\ observations (with the G230L grating) were usually contemporaneous to within a day of those of FUV lines (with the G130M and G160M gratings), though in some instances the two were separated by years.
The dashed line in the middle panel scales the FUV pseudocontinuum flux integration to fill gaps in its 1160-1740~\AA\ range.
See text for details on \lya\ estimation.
The flux of \Mgii\ and \lya\ exceeds FUV lines at all ages, whereas the FUV pseudocontinuum transitions from contributing less to more flux than FUV lines at a period of around 30~d, between Hyades and field ages.
\label{fig:ratioEvol}}
\end{figure}

Emission sources formed at different regions in the stellar atmosphere exhibit different rates of evolution (Figures \ref{fig:ageEvol} - \ref{fig:RoEvol}).
Figure~\ref{fig:indices} compares the posterior distributions of $\alpha$ from the MCMC evolution fits to each source of emission.
As a function of age and rotation, \Mgii\ and FUV pseudocontinuum emission decline less rapidly than most other emission lines.
Though ISM absorption is a factor for \Mgii, it cannot fully explain this difference (Section~\ref{subsec:ISM}).
To quantify the statistical difference in indices, we determined the fraction of MCMC samples for which the \Ciii\ decline is as slow or slower than that of \Mgii\ and the FUV pseudocontinuum.
For \Mgii\ this occurred in \replaced{0.014\%}{0.005\%} of samples (\replaced{$3.8\sigma$}{$4.1\sigma$}) with respect to age and \replaced{0.0023\% of samples}{none of 177,000 samples} (\replaced{$4.2\sigma$}{$>4.5\sigma$}) with respect to rotation period. 
For the FUV pseudocontinuum, it occurred in \replaced{2.6\%}{4.6\%} of samples (\replaced{$2.2\sigma$}{$2.0\sigma$}) with respect to age and \replaced{0.97\%}{1.3\%} of samples (\replaced{$2.6\sigma$}{$2.5\sigma$}) with respect to rotation period.
This difference in rates of decline causes the ratio of \Mgii\ and FUV pseudocontinuum emission to FUV line emission to evolve by on order of magnitude from young to old ages (Figure~\ref{fig:ratioEvol}).

The more gradual evolution of \Mgii\ and (marginally) the FUV pseudocontinuum versus FUV lines must stem from differences in where these lines are formed in the stellar atmosphere.
Whereas the FUV lines we analyzed form almost entirely from plasma in the transition region of the stellar atmosphere, \Mgii\ and FUV pseudocontinuum emission form largely in the stellar chromosphere \citep{avrett08,leenaarts13a,linsky12}.
This supports evidence from previous studies that emission from cooler regions of activity-heated stellar atmospheres evolves more slowly with age.
\cite{guinan03} reported such a trend in \textit{Far Ultraviolet Spectroscopic Explorer} observations of solar analogs, with $\alpha$ ranging from 0.88 (\Cii~1037~\AA) to 1.9 (\ion{Fe}{18}~975~\AA).
For early M stars, \cite{shkolnik14} measured $\alpha=1.36\pm0.32$ in \textit{ROSAT} X-ray flux, $0.99\pm0.19$ in \textit{GALEX} FUV flux, and $0.84\pm0.09$ in \textit{GALEX} NUV flux.
The \textit{GALEX} FUV band's coverage of roughly 450~\AA\ of the FUV pseudocontinuum, while including only three strong emission lines (\Siiv, \Civ, and \Heii), could explain why its decline as measured by \cite{shkolnik14} so closely matches that of the FUV pseudocontinuum we measured.
EUV fluxes predicted from modeling by \cite{peacock20} of early M star upper atmospheres matched to \textit{GALEX} fluxes yield an age-decline of $\alpha=1.0$.

The difference in the evolution of \Mgii, the FUV pseudocontinuum, and FUV lines could indicate physical differences in the evolution of M star chromospheres and transition regions.
\cite{peacock20} simulated stellar EUV-NUV emission by modifying PHOENIX models to include a hot upper atmosphere over the star's photosphere.
They defined the temperature structure of the upper atmosphere with three parameters: the depth (in column mass) of the chromospheric temperature minimum, the depth (again, in column mass) at which the transition region begins, and the temperature gradient (versus pressure) of the transition region.
While all parameters have an effect across the UV, NUV emission is more sensitive to the depth of the chromospheric temperature minimum than FUV emission.
FUV emission is most sensitive to the depth and temperature gradient of the transition region.
The change in the \Mgii\ (chromosphere) to FUV line (transition region) flux ratio in Figure~\ref{fig:ratioEvol} during the declining activity phase suggests that the chromospheric temperature minimum is moving upward more slowly than the start of the transition region, though a change in the slope of the transition region temperature gradient could also contribute.
In turn, these changes could reflect differences in how the chromosphere and transition region are heated in these stars.
If the chromosphere is heated primarily by acoustic shocks launched from photospheric convection rather than magnetic activity \citep{wedemeyer04}, it would be less affected than the transition region as M stars age.

\added{\cite{linsky20} recently conducted a similar analysis, examining the ratio of X-ray (coronal) to reconstructed \lya\ (chromospheric) emission as a function of spectral type and activity level.
They found that X-ray emission decreases more rapidly than \lya\ emission toward lower activity levels (older ages) for all types, akin to the more rapid evolution of FUV line versus \Mgii\ emission we observed.
For M stars, \cite{linsky20} found this discrepancy to be less pronounced, suggesting the differences in line evolution we have observed would grow toward earlier spectral types.
}

The budget of UV emission changes considerably over the lifetime of an early M star.
At around a period of 30~d, between the age of the Hyads and the field stars, the FUV pseudocontinuum begins to account for more FUV flux than strong FUV lines, excluding the \lya\ line, which dominates at all ages (Figure \ref{fig:ratioEvol}).
Assuming the empirical $F_\mathrm{Ly\alpha} = 60 F_\mathrm{Mg\ II}^{0.77}$ relationship of \cite{youngblood16} is constant with age and rotation, \lya\ would evolve as $t^{-0.6}$, $P_\mathrm{rot}^{-0.9}$, and \replaced{$Ro^{-0.8}$}{$Ro^{-0.9}$} (though see Section \ref{subsec:ISM}).
Its emission exceeds combined FUV line and pseudocontinuum emission by a factor of 5-7 during saturation, increasing to 40-80 by 10~Gyr.
These changes in the UV energy budget with age will affect the balance of molecular photolysis in the upper atmospheres of orbiting planets.

\subsection{ISM Absorption of \Mgii}
\label{subsec:ISM}
The more gradual evolution of \Mgii\ flux begs the question of whether ISM absorption could be at play.
Because the younger stars of the sample are further away, the column density of $Mg^{+}$ ions in the line of sight will likely be larger for them.
This could suppress the \Mgii\ flux of those stars, artificially shrinking the difference between the \Mgii\ flux of the young (far away) and old (nearby) stars (i.e., lowering $\alpha$).

ISM absorption is likely affecting the \Mgii\ fluxes we measured.
For cool, main sequence stars at distances representative of the Tuc-hor objects (35-50~pc), \cite{redfield02} measured \Mgii\ column depths between $10^{12.34}$ and $10^{13.14}$~\percmsq\ and Doppler broadenings of 2.1 to 4.3~km~\pers.
For a star with an intrinsically narrow \Mgii\ line such as that of GJ~832 (FWHM 18~km~\pers; \citealt{france13}), these ranges correspond to absorption varying from 17-53\%.
Broader \Mgii\ lines will have more flux shifted outside of the narrow ISM absorption, lessening this effect.
Of the young stars in the sample, the FWHMs of our \Mgii\ fits for 11 of 18 stars  are greater than 50~km~\pers, well above the 30~km~\pers\ LSF of the G230L grating.
For the broadest of these, ISM absorption could fall below 1\%.

This range of possible ISM absorption is too small to explain the more gradual evolution of \Mgii\ emission with age.
However, it could contribute to the scatter of \Mgii\ emission (Figure~\ref{fig:PEvol}) and its apparent evolution with rotation period even during saturation for the Tuc-Hor objects (Figures \ref{fig:PEvol} and \ref{fig:ratioEvol}).
The Tuc-Hor objects span an order of magnitude in rotation period, with rotational speeds fast enough to significantly broaden their emission lines.
The broader lines of faster rotating stars will be less affected by the narrow ISM absorption, potentially producing the observed trend.
However, statistically this trend is marginal, with a $p$-value of 0.005 (2.6$\sigma$) that weakens to 0.02 (1.7$\sigma$) when considering each panel of Figure~\ref{fig:PEvol} as a separate trial of the null hypothesis.

If ISM absorption is impacting the observed flux of the slowest rotating Tuc-Hor objects, then the \Mgii\ emission of the fastest rotating objects could be a better estimate of the \Mgii\ saturation level.
Accordingly, the rates of \Mgii\ decline with age, rotation, and Rossby number should be regarded as lower limits, with the true $\alpha$ values possibly up to 0.2 higher.

\subsection{Predicting the UV Emission of M Stars, Present, Past, and Future}
\label{subsec:predict}
\replaced{The relationships we have derived add}{This work adds} to a growing list of resources readers could use to estimate the UV emission of early M stars.
Such resources are invaluable given that \textit{HST}'s eventual end of life is likely to leave a long gap in UV observational capabilities.
For convenience, Table \ref{tbl:recipe} outlines and compares the resources presently available.

\added{\subsubsection{UV Line Fluxes}}
The relationships we derived enable the emission from strong UV lines and the FUV pseudocontinuum to be estimated from age or rotation period with a $1\sigma$ precision of 0.2~dex, less than a factor of 2.
There is greater scatter in the fits to Rossby number than age or rotation period, with a typical precision of 0.2-0.3~dex.
For a sample of stars diverse in spectral type, casting quiescent activity evolution in terms of Rossby number should, by design, increase the consistency of points around a saturation-decay model by accounting for differences in the strength of convection.
In this case, however, the sample spans a narrow range in spectral type, and it appears that attempting to infer convective turnover rates in an effort to improve quiescent activity evolution fits only introduces noise, perhaps due to metallicity differences in the field sample in particular (Table \ref{tbl:starprops}).
We include the Rossby number fits due to their long heritage in the literature and because they could be of use in estimating UV fluxes for a star outside of the M0 to M2.5 range.
However, because our fits exclude any such stars, doing so would essentially be an extrapolation and should accordingly be done with caution.
The UV line flux measurements versus rotation period plots of \cite{france18} provide a useful point of comparison for a broad range of spectral types.

\begin{deluxetable*}{llp{.15\textwidth}p{.15\textwidth}p{.2\textwidth}p{.3\textwidth}}
\tablewidth{\textwidth}
\tabletypesize{\footnotesize}
\caption{Options for predicting the \replaced{FUV}{UV} emission of early M stars. \label{tbl:recipe}}
\tablehead{\colhead{} & \colhead{Stellar Type} & \colhead{Data Available} & \colhead{Emission Source} & \colhead{Recommended Procedure} & \colhead{Accuracy, Limitations, and Caveats}\\
\colhead{} & \colhead{} & \colhead{} & \colhead{of Interest} & \colhead{} & \colhead{}}
\startdata
1 & M0 - M2.5 & rotation period or age & \Ciii, \Siiii, \Nv, \Cii, \Siiv, \Civ, \Heii, \Mgii\ or unresolved FUV pseudocontinuum & scale using relationships in Table \ref{tbl:evol} & $<$ a factor of 2 \\
2 & M1 and M2 & rotation period or age & full UV spectrum, resolved continuum and EUV & pick closest match from \cite{peacock20}, scale by \Prot$^{-2}$ or $t^{-1.3}$ & factor 1-10 in major FUV lines, more accurate in broadband flux (see \citealt{peacock19b}), some strong spurious lines can be identified and masked by comparison to a spectrum from this work or \cite{loyd16}, does not include coronal EUV emission sources (a minimal contributor to total EUV flux), consider photosphere when scaling above 2000~\AA\ (Section \ref{subsec:predict}) \\
3 & K1 - M5 & rotation period or age & full UV spectrum, unresolved continuum and EUV & pick closest match from \cite{loyd16} (reference also \citealt{france16,youngblood16}), scale by \Prot$^{-2}$ or $t^{-1.3}$ & factor of 2-4 in major FUV lines and broad continuum, \Oi~1305~\AA\ lines masked due to telluric contamination, no data below roughly 1150~\AA, consider photosphere when scaling above 2000~\AA\ (Section \ref{subsec:predict}) \\
4 & M0 - M9 & \Caii\ $S$ index, \Caii\ $R'_{HK}$, or $L_\mathrm{H\alpha}/L_\mathrm{bol}$ & \Siiii, \lya\, \Nv, \Siii, \Cii, \Siiv, \Civ, \Heii, \Mgii & \cite{melbourne20} & factor of 2-4 if using  $S$ or $R'_{HK}$ or 2.5-5 if using $L_\mathrm{H\alpha}/L_\mathrm{bol}$ \\
5 & K-M & \Mgii, \Siiii, or \Civ\ flux & \lya\ & \cite{youngblood16} & generally 10\%, unknown \lya\ self-absorption could introduce factor 2 inaccuracy\\
6 & M1 and M2 or K1-M5 & \textit{GALEX} FUV or NUV or other UV emission measurement & full UV spectrum & see (2) or (3), subtract photosphere, scale remainder to match measured flux, add photosphere & see (3), beware that scaling by a single emission line could yield corresponding inaccuracy in broadband FUV flux, consider photosphere when scaling above 2000~\AA\ (Section \ref{subsec:predict})\\
7 & M0 - M2.5 & None & evolution of \Ciii, \Siiii, \Nv, \Cii, \Siiv, \Civ, \Heii, \Mgii\ or unresolved FUV pseudocontinuum & see (1) & $<$ a factor of 2 for the population average; however, saturation lifetime could vary by more than an order of magnitude for individual stars \citep{tu15} and the rate of spin-down and associated activity decline is not likely to be constant (Section \ref{subsec:predict})\\
8 & M1 and M2 & None & evolution of full UV spectrum, resolved continuum and EUV & interpolate from age grid of \cite{peacock20} & factor 1-10 in major FUV lines, more accurate in broadband flux, see also (2) and (8)
\enddata

\end{deluxetable*}

The precision with which rotation period and age predict stellar FUV line emission for early M stars exceeds that of optical activity indicators.
Recently \cite{melbourne20} assembled a sample of nearly 70 stars spanning M0 to M9 types and computed empirical fits of UV line fluxes versus \Caii\ $S$-index, \Caii~\Rhk, and $L_\mathrm{H\alpha}/L_\mathrm{bol}$ for the sample, updating earlier work by \cite{youngblood17}.
Scatter about their power-law fits between \Caii~\Rhk\ and UV lines range from 0.4-0.6~dex, with the exception of \Mgii\ at 0.31 dex.
Their sample included all HAZMAT targets and covered ages ranging from 10~Myr to 10~Gyr.
The larger scatter in these UV-optical relationships versus the UV age and rotation relationships we derived could come from a number of sources.
Chief among these are the larger range of spectral subtypes and the gaps in time between the optical and UV activity observations.
Flares, though sometimes present, likely do not contribute significantly \citep{melbourne20}.
Differences in the formation region of \Caii~H~\&~K emission versus strong FUV lines could be a factor, since \Caii~H~\&~K emission forms primarily in the stellar chromosphere \citep{bjorgen18}.
This is supported by the lower level of scatter in fits between \Caii~H~\&~K and \Mgii\ than \Caii~H~\&~K and FUV lines.

\added{\subsubsection{Scaling Spectra}}
The entire FUV spectrum (with the exception of \lya) of an early M star can be predicted reasonably well by simply scaling the FUV spectrum of a template star \replaced{by $P_\mathrm{rot}^{-2}$}{along a saturation-decline track}.
The bottom panel of Figure~\ref{fig:spectra} illustrates this, wherein all spectra have been scaled to match a rotation period of 10~d via a saturation-decline track with $t_\mathrm{sat}=10$~d and $\alpha=2$. 
This brought the spectra into good alignment with each other, differing in both the line and continuum regions generally by less than a factor of a few for the FUV.

The NUV regions of the spectra in Figure~\ref{fig:spectra} (the data spanning 2750-3250~\AA) show a markedly different trend than the FUV.
NUV pseudocontinuum emission at 650~Myr and Gyr ages is almost identical, but at 40~Myr it is elevated by an order of magnitude.
Meanwhile, FUV pseudocontinuum emission differs by an order of magnitude between each age group.
The reason for this is simply the differing contribution of emission from the stellar photosphere, which shows comparatively little evolution with age.
For field-age stars, NUV pseudocontinuum emission is dominated by the stellar photosphere (e.g., \citealt{loyd16}, Figure~12).
For 40~Myr stars, activity-generated emission dominates, elevating the NUV pseudocontinuum, but by 650~Myr activity-generated emission has already fallen below photospheric levels.
A similar result is evident in the analysis of
\cite{shkolnik14}.
Their work showed that the fraction of emission attributable to early M star photospheres in the \textit{GALEX} NUV bandpass, while minimal at young ages, can reach 40\% by Gyr ages.
The \textit{GALEX} NUV band (predominantly 1800-2750~\AA) lies blueward of the region under consideration here (2750-3250~\AA), so it is not surprising that photospheric emission dominates as early as 650~Myr onward in the present case.
This fact is what causes the NUV portions of the spectra in the lower panel of Figure \ref{fig:spectra} to reverse order from the top panel.

\added{\subsubsection{Caveats and Areas for Improvement in Predicting UV Emission}}
The ages of the field stars are very uncertain, influencing the fits.
Incorporating these uncertainties into the evolution fits described in Section \ref{subsec:evolfits} increased uncertainty in the rate of activity decline by up to 50\% and allowed the fits to better accommodate the Hyades data.
However, the fits assumed these estimates and associated uncertainties are independent.
In reality, the ages could be systematically high or low if the isochrones or gyrochrones used in the age determination are inaccurate (see, e.g., \citealt{angus20} for an investigation of systematic errors in M star gyrochrones).
The possibility of a systematic offset means that the uncertainty in the fits to activity evolution with age should be taken as a lower limit.
A systematic shift of ages to higher or lower values in log space would correspond to a roughly equivalent change in the value for the index of the power law decline, $\alpha$.
We expect any systematic offset in the ages of the field stars to be no more than a factor of two (0.3~dex), implying any additional uncertainty in $\alpha$ for the age fits of Table \ref{tbl:evol} will be under 0.3.
Since this would affect all lines equally, it does not affect our conclusions regarding differences in evolution between emission sources.

The differences between the model and photometric radii of young stars is a factor in predicting their UV luminosity.
The 10-60\% differences between these two radii correspond to a factor of 1.2-2.6 in surface flux.
This means that scaling a surface flux from this work to the model radius of a pre-main-sequence star could yield an underestimate of the star's UV luminosity.

An important factor in fitting the quiescent activity evolution of a sample of stars is that this fit will represent a population average.
In reality, each star will follow its own track, equivalent to an ensemble of lines instead of a single line in Figure~\ref{fig:ageEvol}.
The primary difference in this ensemble of tracks would be the lifetime of saturation.
As a result, the scatter in quiescent activity levels for a group of coeval stars will grow as stars in that group progressively transition from saturated to declining activity.

Evidence is mounting that the spin down rate of M stars varies with time.
The spin down of K through early M stars appears to stall around the age of the Hyades, with the spin down of lower-mass objects stalling for a longer period of time  \citep{agueros18,curtis19,douglas19}.
Early M stars in the \textit{Kepler} field make up two distinct populations of rotators, with the modal rotation period of one population at 19~d and the other at 33~d \citep{mcquillan13}.
This could be a result of episodic star formation, as \cite{mcquillan13} infer, or stalled spin down around 19~d.
At lower masses, \cite{newton16,newton18} found a wide range of rotation periods, 10-70~d, that are poorly represented among the population of nearby mid M stars and concluded that this is the result of periods of slowed and accelerated spin down.

\deleted{More work on the rotational evolution of Ks and Ms is needed to produce a consistent and comprehensive picture of the spin down, and, consequently, activity evolution of early M stars.}
Modeling by \cite{spada20} that incorporates rotational coupling between the radiative and convective zones of stars yields variable spin-down rates and shows promise in reproducing the age-rotation demographics of low-mass stars \citep{angus20}.
\added{Coupling a spin-down model like that of \cite{spada20} to activity-rotation relationships like those presented here can yield a model where} each star follows its own track based on its initial rotation rate, and where these tracks do not decline linearly in log space but instead reflect variable spin down rates.
\added{\cite{johnstone20b} recently created a coupled activity-spin-down model for the purposes of predicting the combined EUV and X-ray (together, XUV) evolution of stars that powers mass loss from orbiting planets.
Future work could do the same for FUV emission to enable better modeling of photochemistry in planetary atmospheres as stars evolve.
As data on M star rotational evolution improves, coupled age-rotation-activity models will become increasingly accurate.}

\deleted{In time, it might become clear that a more accurate model to fit to Figure~\ref{fig:ageEvol} is one in which
Regardless of how early M stars spin down, their activity and spin are intimately linked.}

\subsection{What are the sources of scatter in these relationships?}
\label{subsec:scatter}
The scatter in the quiescent activity of stars of the same age in this sample is intrinsic to the star.
Measurement uncertainties are well below the 0.2-0.3~dex scatter.
The absolute flux calibration of COS is accurate to roughly 5\% (0.02~dex; \citealt{ihb}).
Two other potential sources of scatter are immediately dismissable.
Flares cannot be a significant contributor to the scatter because these data are time resolved, which enabled us to clean the spectra of detectable flares \citep{loyd18b}.
Note, however, that flares are an important, possibly dominant, source of both line and continuum UV emission in M stars \citep{loyd18a,loyd18b}.
Differences in spectral subtypes cannot be the source of scatter because there is no clear ordering by subtype (point size in Figures \ref{fig:ageEvol}, \ref{fig:PEvol}, and \ref{fig:RoEvol}) in whether line fluxes lie above or below the evolution trend line.

Eliminating the above sources of scatter leaves metallicity differences, scatter in seed rotation rates, age uncertainty, long-term activity cycles, and rotational variability as possible sources of the observed scatter in quiescent stellar activity.
For the Tuc-Hor sample, metallicity, seed rotation rates, and age uncertainty cannot contribute to the observed scatter.
Differences in their seed rotation rates cannot contribute because the stars are all in the saturated activity regime, where rotation rate does not influence quiescent activity.
They are unlikely to have substantial differences in metallicity or age because they formed from the same cloud \citep{tabernero12}.
Hence, the scatter of the Tuc-Hor sample can only be due to long-term activity cycles or rotational variability.
These two effects, combined, must account for the roughly 0.1~dex (25\%) scatter in the fluxes of the FUV emission lines from these stars.
\added{This is supported by a recent analysis of X-ray emission from stellar populations by \cite{johnstone20b} that controlled for various sources of scatter and found activity cycles to be the dominant contributor, though the scatter they analyzed was at the level of 0.36~dex.}

The only difference between the Tuc-Hor and Hyades groups that could influence differences in their scatter is variability in their seed rotation rates.
Because the Hyades stars are (mostly) no longer in the saturated activity regime, differences in their seed rotation will affect their activity evolution by setting the point at which the star will transition from saturated to declining activity.
For Sun-like stars, differences in seed rotation can drive differences in the duration of their saturated activity spanning over an order of magnitude \citep{tu15}.
The present sample reflects this effect, with scatter growing from 0.1~dex at the age of Tuc-Hor to $>0.3$~dex ($>100$\%) at the age of the Hyades.
LP5-282 drives this scatter; however, it is not a rotational outlier when compared to a larger group of early M Hyads \citep{douglas19}.
Within the present sample, LP5-282 likely formed with the highest seed rotation rate, delaying its transition out of saturated activity and shifting its evolutionary track as a function of age to the right by about a factor of two in Figure \ref{fig:ageEvol}.

For the field sample, differences in metallicity and age uncertainty are added as possible contributors to scatter.
This likely explains why this group has the largest scatter of the three.

\subsection{Does the First 100~Myr of Emission Matter Most to Planets?}
\label{subsec:king}
Recently, \cite{king20} challenged the widespread conclusion that the first 100~Myr of a star's \replaced{X-ray and EUV (together, XUV)}{XUV} emission are responsible for the bulk of planetary atmospheric loss (e.g., \citealt{lopez13,luger15,owen17}).
There are no direct measurements of the age-evolution of the ``soft'' (360-912~\AA) range of stellar EUV (100-912~\AA) evolution because absorption from intervening ISM gas blocks  it from observation.
\cite{king20} inferred the evolution of stellar EUV emission by combining observations of the age-evolution of stellar X-ray emission \citep{jackson12} with an empirical trend for the Sun that relates X-ray to EUV luminosity variations \citep{king18}.
The resulting slope in the power-law decay of EUV emission, $\alpha$,  spans 0.63-0.81 and, for soft EUV emission, 0.26-0.33, for FGK spectral types.
For the lowest mass bin of stars, with $B-V$ from 1.275 to 1.41, corresponding to roughly K6 to M0 stars\footnote{From the table compiled by STScI, \url{https://www.stsci.edu/~inr/intrins.html}},  $\alpha=0.68\pm0.22$ (full EUV) and $0.28^{+0.20}_{-0.09}$ (soft EUV).
Such shallow rates of decline would make the EUV emission of a star at Gyr ages more important to planetary atmospheric escape than previously thought.

The actual rate of decline in EUV emission from late K and early M stars, and probably other types as well, is likely to be more rapid than \cite{king20} suggest.
An M star's EUV emission comes primarily from transition region plasma, albeit with significant contributions from the corona and chromosphere \citep{fontenla16, peacock19b}.
It therefore seems likely that the age-evolution of broadband EUV emission will fall near that of FUV emission from the transition-region, roughly $\alpha=1.3$, and within the bounds set by the evolution of chromospheric \Mgii\ and coronal X-ray emission, $\alpha = 0.8 - 1.35$ \citep{shkolnik14}.
\cite{peacock20} derive a value in this range, $\alpha=1.0$, for the evolution of early M star EUV emission based on stellar atmospheric models anchored by broadband FUV and NUV flux measurements from \textit{GALEX}\added{, excluding coronal EUV emission.}
In contrast, taking the late-K/early-M bin from \cite{king20} to be comparable to our sample, the \cite{king20} values for $\alpha$ imply that EUV emission declines more slowly than emission both redward and blueward of the EUV band for early M stars.
We caution, however, that the overlap in the spectral types of these two samples is marginal, intersecting only at M0.

\begin{figure}
\includegraphics{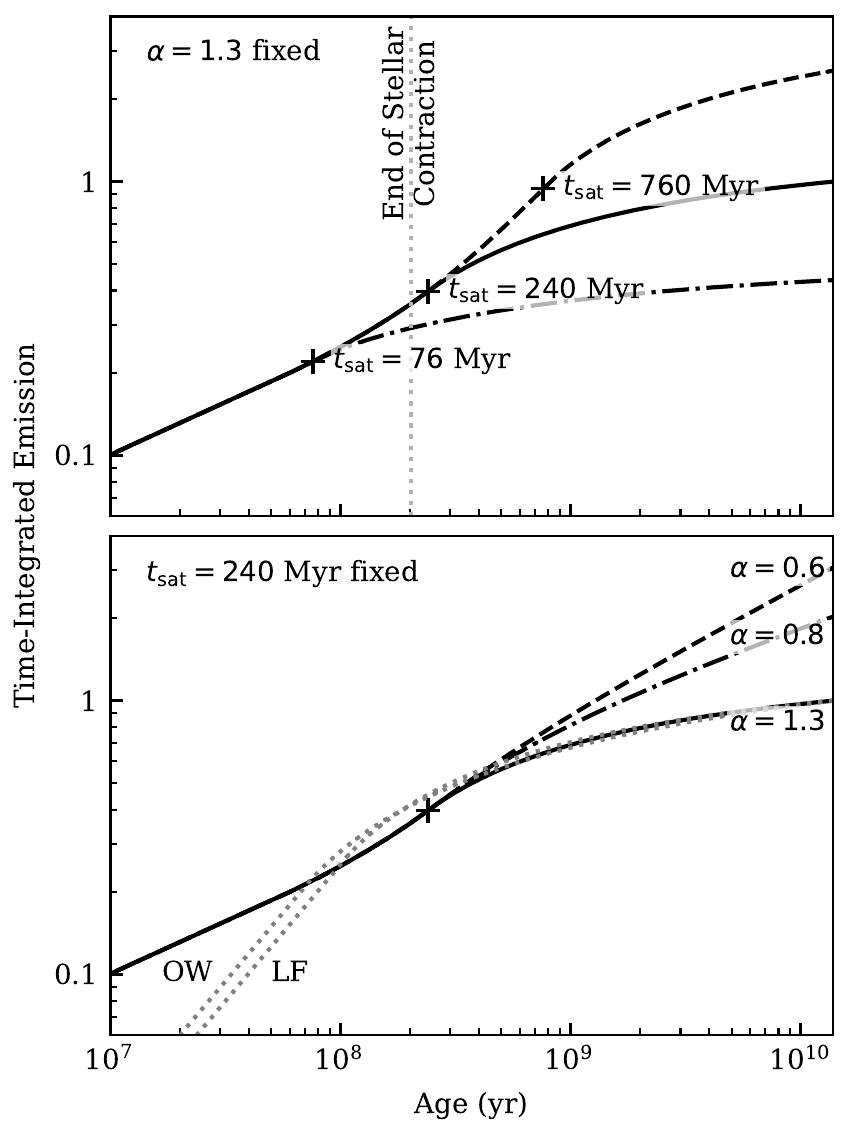}
\caption{Cumulative emission for a variety of activity evolution scenarios.
The upper plot varies the activity saturation lifetime to show the possible effect of different stellar seed rotation rates.
The lower plot shows the effect of different indices of the power law decline.
A power-law index of $\alpha=1.3$ is typical of FUV lines and X-ray emission, 0.8 matches \Mgii, and 0.6 is a possible value for \lya\ (see text).
Note that all curves are normalized to the value at 13.8~Gyr for $t_\mathrm{sat}$ = 240~Myr (our best fit value) and $\alpha = 1.3$.
The pre-main-sequence stellar contraction of a 0.5~\Msun\ star is included, ending at the dotted vertical line.
These curves likely bracket cumulative EUV emission, though we consider the $\alpha=1.3$ curve (valid for FUV line emission) as likely to be the best proxy for the EUV.
\added{Dotted lines in the lower plot show the models adopted by \citet[LF]{lopez13} and \citet[OW]{owen13}, normalized to the value at 13.8~Gyr.}
\deleted{Regardless, }The majority of energy is emitted after 100~Myr \added{in most cases}.
\label{fig:cumulative}}
\end{figure}

Regardless of the true rate of decline in EUV emission, \cite{king20}'s conclusion that the majority of XUV radiation is emitted on Gyr timescales holds true.
Figure~\ref{fig:cumulative} plots the time-integral of emission over the age of the universe for saturation lifetimes spanning an order of magnitude \citep{tu15}, centered on our best-fit value of $t_\mathrm{sat}=240$~Myr, and rates of decline ranging between the value we estimate for \lya, $\alpha=0.6$, to that we found to be typical of our fits to transition-region FUV line emission, $\alpha=1.3$.
We accounted for changes in stellar radius of a 0.5~\Msun\ star as predicted by the MIST stellar evolution models \citep{dotter16,choi16}, though see Section \ref{sec:obs}.
Lifetime-integrated emission varies by a factor of a few between these scenarios.
Mathematically, a rate of decline with $\alpha>1$ implies that the cumulative flux will approach an asymptote with age, whereas $\alpha<1$ implies cumulative flux is unbounded.

In \replaced{all}{most} cases, the majority of lifetime-integrated emission occurs after 100~Myr.
\replaced{This conclusion would not change unless $t_\mathrm{sat}$ falls below 40~Myr for $\alpha=1.3$.}{The exception is the $\alpha=1.3$ and $t_\mathrm{sat}= 76$~Myr case, for which the cumulative emission is roughly evenly divided before and after the 100~Myr mark.}
\added{For comparison, we also plot the time-integral of the XUV emission tracks adopted by \cite{lopez13} and \cite{owen13}, works that predicted the later observed exoplanet radius valley.
These XUV emission tracks are from \cite{ribas05} and \cite{jackson12} for G stars.
Note that they are normalized to their values at 13.8~Gyr for comparison, even though G star planets will, in general, receive less XUV flux over their lifetimes than M star planets with similar equilibrium temperatures.}

Although the bulk of XUV emission might occur \replaced{on Gyr timescales}{after 100~Myr}, the first 100 Myr or so of a star's XUV emission could still be the most important to determining a planet's atmospheric loss.
For a planet whose mass is dominated by its core, mass is most easily removed early in life when the atmosphere is in its most distended state, lofting gas higher out of the planet's gravitational well.
As both mass is lost and the atmosphere radiatively cools and contracts, the remaining mass recedes deeper into the planet's gravitational well.
Hence, atmospheric loss would slow with time even if XUV irradiation did not decline at all (e.g., \citealt{luger15,owen17}).
This could be exacerbated by a transition from recombination-limited to energy-limited mass loss \citep{murray09}, causing further drops in XUV irradiation to drive a linear ($\Dot{M} \propto F_\mathrm{XUV}$) instead of a square root ($\Dot{M} \propto F_\mathrm{XUV}^{0.5}$) drop in the mass loss rate.
We expect that the combination of these considerations will make the first 100~Myr a critical phase of atmospheric loss for a planet, regardless of whether XUV emission declines as $t^{-0.7}$, as \cite{king20} propose, or at a rate closer to $t^{-1.3}$, as we propose.

\section{Summary}
\label{sec:summary}

We have presented results from \textit{HST} UV spectroscopy of 25 M0-M2.5 stars in three distinct age groups, 40~Myr, 650~Myr, and field ages.
For the first time, we resolved the evolution of individual \added{UV} emission lines with both age and rotation.
We found that the surface flux of \Mgii, an oft-used proxy for \lya, as well as surface flux within broad FUV pseudocontinuum bands, declines by 1.3 and 1.7 orders of magnitude, respectively, from 40~Myr to 10~Gyr, whereas strong FUV lines drop by roughly 2.1 orders of magnitude.
These differences are in line with previous observations showing that emission formed lower in the activity-heated upper atmospheres of stars declines  less rapidly with age.
In consequence, the energy budget of UV flux from early M stars becomes weighted more toward ``cooler'' sources of emission as early M stars age.

We fit a model of constant saturation followed by a power-law decay to the evolution of each emission line.
Equipped with an age or rotation period for an arbitrary early M star, the reader can use fits from Table~\ref{tbl:evol} to predict that star's quiescent emission in the strong UV lines \Cii, \Ciii, \Civ, \Heii, \Mgii, \Nv, \Siiii, and \Siiv\ and the unresolved FUV pseudocontinuum to a $1\sigma$ accuracy of 0.2-0.3~dex.
This is the most precise means presently available to predict these sources of emission in lieu of observation.
The youngest stars exhibit 0.1~dex scatter, the source of which can be confined to long-term activity cycles, rotational modulation, or an unknown source of variability.
We compiled a quick-reference list of available resources for predicting the unobserved UV emission of a star and their distinguishing characteristics in Table \ref{tbl:recipe}.

The fits to line evolution with age we have established can be used in models of the photochemical state of an exoplanetary atmosphere at various stellar ages and its evolution through time.
Further, they provide a likely bound on the EUV emission of early M stars.
Because both the FUV line and X-ray emission of early M stars declines as $t^{-1.3}$, we suggest that EUV emission declines at a similar rate.
This is steeper than the rate of decline inferred by \cite{king20} for K6-M0 stars, but does not alter their conclusion that the bulk of EUV emission occurs after 100~Myr.
However, other considerations are likely to make the first 100~Myr of an early M star's life, when activity is saturated and planetary atmospheres are at their most distended, a critical phase in the atmospheric evolution of orbiting planets.
Readers employing our age-evolution fits should bear in mind that order-of-magnitude differences in stellar seed rotation rates can lead to corresponding differences in activity saturation lifetime between individual stars \citep{tu15}.

\acknowledgments
Support for program 14784 was provided by NASA through a grant from the Space Telescope Science Institute (HST-GO-14784.001-A), which is operated by the Association of Universities for Research in Astronomy, Inc., under NASA contract NAS 5–26555.
R.O.P.L. and E.S. thank STScI and NASA for this support.
We acknowledge with thanks the variable star observations from the AAVSO International Database contributed by observers worldwide and used in this research. 
The team at STScI were of great help in planning these observations, and we thank them for their responsiveness and attention to detail.
Stephanie Douglas and Jason Curtis assisted with determining stellar parameters, including rotation periods and effective temperatures, of the Hyads in the sample, for which we are very grateful.
We also wish to thank several scientists who went above and beyond to provide us with extended or more detailed versions of published data products, including George McDonald, who provided posterior distributions for saturation-decay fits;  Adam Kraus, who provided optical spectra for possible binary J22025; and  Federico Spada, who provided a complete set of stellar gyrochrones.
We are grateful to Ruth Angus, Melodie Kao, Tahina Ramiaramanantsoa, Wilson Cauley, and Ella Osby for helpful advice, questions, and conversations.

This research is based on observations made with the NASA/ESA \textit{Hubble Space Telescope}, obtained from the Data Archive at the Space Telescope Science Institute. These observations are associated with program \# 14784, with additional data from programs \# 12464, 13650, 14767, and 15174. This paper includes data collected with the \textit{TESS} and \textit{Kepler} missions, obtained from the MAST data archive at the Space Telescope Science Institute (STScI). Funding for the TESS mission is provided by the NASA Explorer Program. Funding for the Kepler mission is provided by the NASA Science Mission Directorate. STScI is operated by the Association of Universities for Research in Astronomy, Inc., under NASA contract NAS 5–26555.

\facilities{HST (COS), MAST, K2, TESS, AAVSO, ELODIE, SDSS}

\software{isochrone \citep{morton15}, astropy \citep{astropy13}, stardate \citep{angus19a}, emcee \citep{foreman19}, starspot (\url{https://github.com/RuthAngus/starspot})}

\vspace{6 in}

\bibliography{refs,refs_other}{}
\bibliographystyle{aasjournal}

\appendix

\section{Custom Coaddition of Spectra}
\label{app:coadd}
To coadd spectra obtained using different central wavelength settings, we first established a common wavelength grid covering all wavelengths where at least one spectrum had data at the same resolution as the component spectra.
We then rebinned each spectrum onto this wavelength grid ensuring that total flux was preserved.
Once rebinned, we averaged fluxes and errors (in quadrature), weighting by exposure time.
We masked data in pixels flagged with data quality issues marked as serious within the x1d file headers.
If this resulted in no data in a pixel, then we retained the flagged data.
We propagated all data quality flags for masking in later fits.

\begin{figure*}
\includegraphics{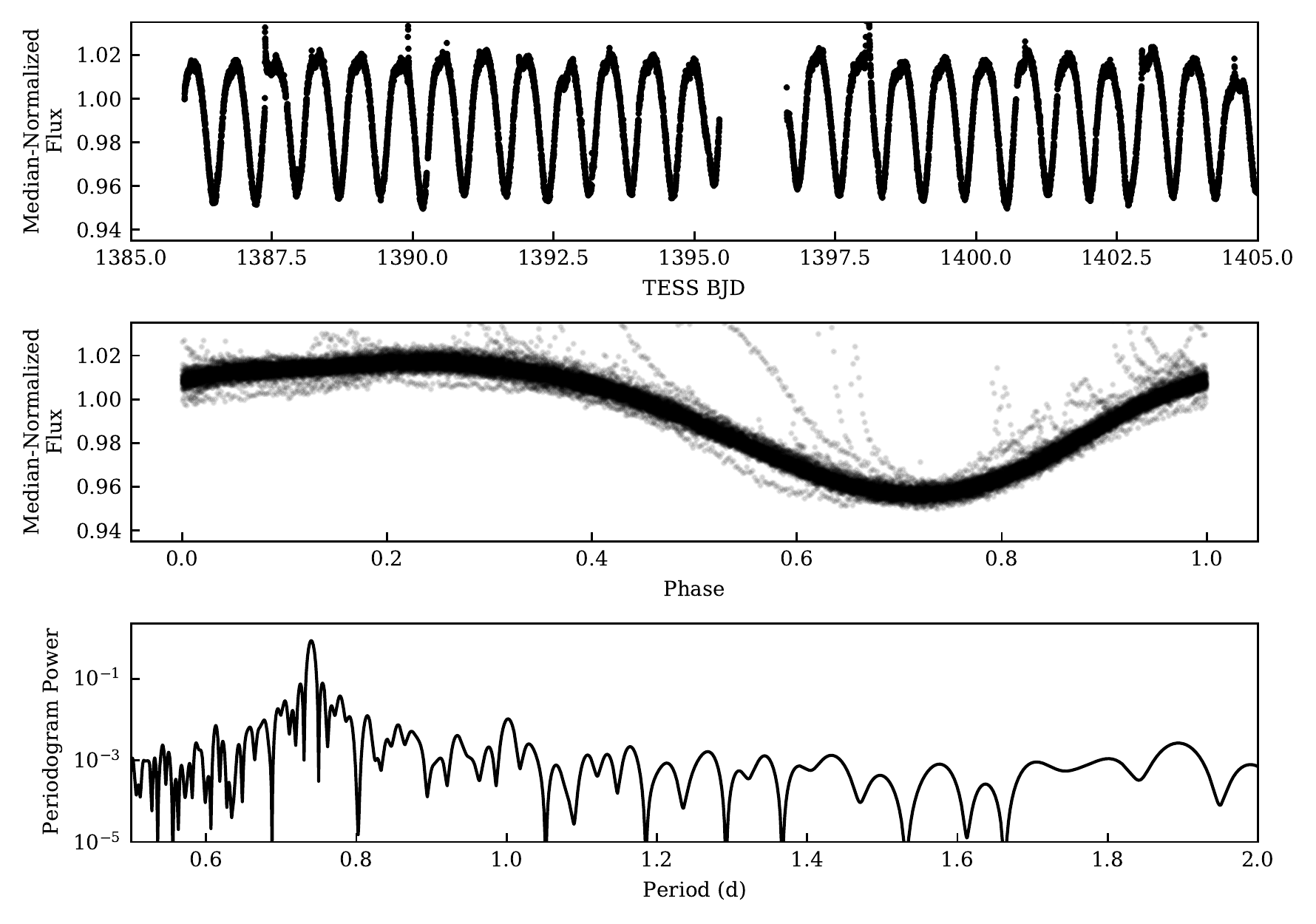}
\caption{\added{\textit{TESS} lightcurve and Lomb-Scargle periodogram for J02365.
The top panel shows one sector of the TESS data.
The middle panel shows the data from both sectors folded on the period of maximum power.
Positive excursions, some off the plot range, are stellar flares.
The bottom panel shows the results of the periodogram analysis with a strong peak at 0.74~d.}
\label{fig:period}}
\end{figure*}

\added{
\section{Periodogram for J02365}
\label{app:period}
Because several rotation periods have been reported in the literature for J02365, we here provide a diagnostic figure for our periodogram analysis (Figure \ref{fig:period}).
This may also serve as an example for other stars for which we derived rotation periods from \textit{TESS} data.
}

\end{document}